\documentclass[a4paper,fleqn,usenatbib]{mnras}

\usepackage{mathptmx}

\usepackage[T1]{fontenc}
\usepackage{ae,aecompl}

\usepackage{graphicx}	
\usepackage{amsmath}	
\usepackage{amssymb}	
\usepackage{color}

\title[Spectropolarimetry of supernovae 2013ej and 2017ahn]{Evidence for multiple origins of fast declining Type~II supernovae from spectropolarimetry of SN~2013ej and SN~2017ahn\thanks{Based on observations made with ESO Telescopes at the La Silla Paranal Observatory under Program IDs 091.D-0401 and 098.D-0852.
}}

\author[T. Nagao et al.]{T. Nagao,$^{1,2}$\thanks{E-mail: takashi.nagao@utu.fi}
F. Patat,$^{2}$
S. Taubenberger,$^{3}$
D. Baade,$^{2}$
T. Faran,$^{4}$
A. Cikota,$^{5,6}$
\newauthor
D. J. Sand,$^{7}$
M. Bulla,$^{8}$
H. Kuncarayakti,$^{1,9}$
J. R. Maund,$^{10}$
L. Tartaglia,$^{11}$
\newauthor
S. Valenti,$^{12}$
and D. E. Reichart$^{13}$
\\
$^{1}$ Department of Physics and Astronomy, University of Turku, FI-20014 Turku, Finland\\
$^{2}$ European Southern Observatory, Karl-Schwarzschild-Str. 2, 85748 Garching b. M\"{u}nchen, Germany\\
$^{3}$Max-Planck-Institut f\"{u}r Astrophysik, Karl-Schwarzschild-Str 1, 85748 Garching b. M\"{u}nchen, Germany\\
$^{4}$Racah Institute of Physics, The Hebrew University of Jerusalem, Jerusalem 91904, Israel\\
$^{5}$ European Southern Observatory, Alonso de Cordova 3107, Vitacura, Casilla 19001, Santiago de Chile, Chile\\
$^{6}$Physics Division, Lawrence Berkeley National Laboratory, 1 Cyclotron Road, Berkeley, CA 94720, USA\\
$^{7}$Department of Astronomy and Steward Observatory, University of Arizona, 933 N. Cherry Avenue, Tucson, AZ 85719, USA\\
$^{8}$The Oskar Klein Centre, Department of Astronomy, Stockholm University, AlbaNova, SE-106 91 Stockholm, Sweden\\
$^{9}$ Finnish Centre for Astronomy with ESO (FINCA), FI-20014 University of Turku, Finland\\
$^{10}$ Department of Physics and Astronomy, University of Sheffield, Hicks Building, Hounsfield Road, Sheffield S3 7RH, UK\\
$^{11}$INAF - Osservatorio Astronomico di Padova, Vicolo dell’Osservatorio 5, I-35122 Padova, Italy\\
$^{12}$Department of Physics, University of California, Davis, CA 95616, USA\\
$^{13}$Department of Physics and Astronomy, University of North Carolina at Chapel Hill, Chapel Hill, NC 27599, USA
}

\date{Accepted XXX. Received YYY; in original form ZZZ}

\pubyear{2021}

\begin{document}
\label{firstpage}
\pagerange{\pageref{firstpage}--\pageref{lastpage}}
\maketitle

\begin{abstract}
The origin of the diverse light-curve shapes of Type~II supernovae (SNe), and whether they come from similar or distinct progenitors, has been actively discussed for decades. Here we report spectropolarimetry of two fast declining Type~II (Type~IIL) SNe: SN~2013ej and SN~2017ahn. SN~2013ej exhibited high continuum polarization from  very soon after the explosion to the radioactive tail phase with time-variable polarization angles. The origin of this polarimetric behavior can be interpreted as the combination of two different aspherical structures, namely an aspherical interaction of the SN ejecta with circumstellar matter (CSM) and an inherently aspherical explosion. Aspherical explosions are a common feature of slowly declining Type~II (Type~IIP) SNe. By contrast, SN~2017ahn showed low polarization not only in the photospheric phase but also in the radioactive tail phase. This low polarization in the tail phase, which has never before been observed in other Type~IIP/L SNe, suggests that the explosion of SN~2017ahn was nearly spherical. These observations imply that Type~IIL SNe have, at least, two different origins: they result from stars that have different explosion properties and/or different mass-loss processes. This fact might indicate that 13ej-like Type~IIL SNe originate from a similar progenitor to those of Type~IIP SNe accompanied by an aspherical CSM interaction, while 17ahn-like Type~IIL SNe come from a more massive progenitor with less hydrogen in its envelope.
\end{abstract}




\section{Introduction}

Type~II supernovae (SNe), which are core-collapse explosions of massive stars, have been historically divided into Type~IIP (plateau) and Type~IIL (linear) according to their light-curve (LC) shapes \citep[e.g.,][]{Barbon1979,Filippenko1997}. Type~II SNe show spectra with continuum radiation superimposed by P~Cygni lines until $\sim100$~d after the explosion (the photospheric phase), and spectra dominated by emission lines (the tail phase) after a sudden luminosity drop. During the photospheric phase, Type~IIP SNe have constant magnitude in the $R$ bands, while Type~IIL SNe show a rapid luminosity decline. The observed fractions of Type~IIP and IIL in a volume-limited sample are $\sim70$ per cent and $\sim10$ per cent of all Type~II SNe, respectively \citep[][]{Li2011}. It has been suggested that the two types differ also in other observational properties in addition to the LC shapes. For example, Type~IIL SNe have higher peak luminosity \citep[e.g.,][]{Patat1993, Patat1994, Li2011, Anderson2014a, Faran2014a, Faran2014b, Sanders2015, Valenti2016, deJaeger2019}, shorter photospheric phases \citep[e.g.,][]{Gutierrez2017, deJaeger2019}, higher expansion velocities \citep[e.g.,][]{Faran2014b} and less pronounced P~Cygni H$_{\alpha}$ profiles \citep[e.g.,][]{Patat1994, Schlegel1996, Gutierrez2014} than Type~IIP SNe. However, the classification between Type~IIP and IIL SNe is ambiguous. In fact, there have been claims of a continuous transition from Type~IIL to IIP SNe \citep[e.g.,][]{Anderson2014a, Sanders2015, Valenti2016}.

Historically, these different appearances have been interpreted in terms of different amounts of hydrogen in the envelopes of their progenitors (red supergiant stars; RSGs) at the time of explosion owing to different mass-loss histories.  That is, SNe with smaller hydrogen envelopes should show steeper luminosity decline in the photospheric phase and so appear as Type~IIL SNe \citep[e.g.,][]{Barbon1979, Blinnikov1993, Nomoto1995, Morozova2015, Moriya2016}. As a cause of the reduced hydrogen envelope, two main mechanisms have been proposed: stellar wind mass loss from single massive stars and mass loss due to interaction with a companion star. In the former case, the origin of the different appearances of Type~II SNe is attributed to different initial mass and/or metallicity of their progenitor stars. Stars with higher mass and/or higher metallicity will produce more Type~IIL-like SNe because of the stronger stellar-wind-induced mass loss  \citep[e.g.,][]{Heger2003}. However, it has been suggested that this scenario has difficulties to explain the fraction of Type~IIL SNe among all Type~II SNe \citep[e.g.,][]{Heger2003}. In the case of envelope stripping in a binary, the diversity is supposed to be due to different binary properties  \citep[e.g.,][]{Nomoto1995}. Stars in binaries with shorter initial separation and/or higher mass ratio explode in a smaller hydrogen envelope because of the stronger interaction \citep[e.g.,][]{Yoon2017, Ouchi2017}. In this case, the zero-age-main-sequence masses of the progenitor stars are considered to be common for Type~IIP and IIL SNe. Therefore, the underlying explosion physics should be common to both types. We note that, in the Local Group, most massive stars are formed in binary systems and, during the course of their evolution, a large fraction goes through phases of interaction \citep[e.g.,][]{Sana2012}.

Based on radiation hydrodynamics simulations, \citet[][]{Morozova2017} recently demonstrated that explosions of ordinary RSGs differing merely in the mass and distribution of the circumstellar matter (CSM) can reproduce the LC diversity in Type~II SNe (with mass-loss rates of $\sim0.1$–0.2 M$_{\odot}$ yr$^{-1}$ over a few months to years and CSM extents of $\sim2100$–2300 R$_{\odot}$). Furthermore, \citet[][]{Hillier2019} modeled the LCs and optical spectra of Type~II SNe using RSG progenitors that have not only different masses and CSM distributions but also different H-rich envelope masses. From the comparison of the models with observations, they concluded that differences in hydrogen envelope mass alone, which is the classical explanation, cannot quantitatively reproduce the LC diversity, and thus that CSM interaction is necessary.

To explain the presence of major amounts of CSM close to the progenitors, mass-ejection due to binary interaction \citep[e.g.,][]{Chevalier2012, Soker2013, Yoon2017, Ouchi2017} and mass-loss due to stellar instabilities \citep[e.g.,][]{Humphreys1994, Langer1999, Yoon2010, Arnett2011, Quataert2012, Shiode2014, Smith2014a, Woosley2015, Quataert2016, Fuller2017, Ouchi2019}, have been proposed.  By contrast, mass-loss due to a stellar wind is not a viable mechanism because it cannot explain the implied short times in which large amounts of CSM need to be piled up in the vicinity of the progenitors. In the case of binary interaction, the same explanation as for the envelope stripping is applicable. Stars in binaries with shorter initial separation and/or higher mass ratio have denser nearby CSM due to the stronger interaction. Similarly, the zero-age-main-sequence masses of the progenitor stars are supposed to be common for Type~IIP and IIL SNe, and thus the underlying explosion physics should be common to both types.  Losses induced by stellar instabilities seem to apply only to some very specific types of progenitors undergoing major mass-loss episodes at the end of their lives, i.e., shortly before the SN explosion, although the exact nature of these objects is still not clear. Therefore, in this case, the underlying explosions are expected to be different between Type~IIP and IIL SNe.  In fact, there are indications that correlations between the circumstances of the explosion and the LC shapes in Type~II SNe do exist. \citet[][]{Morozova2018} have extended their previous work by calculating the LCs of Type~II SNe using RSGs that have not only different CSM masses and distributions but also different stellar masses and explosion energies.  By deriving the properties of the progenitors and their CSM from observations, they found possible correlations between CSM mass and explosion energy and between CSM radial extent and progenitor mass.

Revealing the explosion geometries of Type~II SNe is key to understanding the origin of their observed diversity. Polarimetry of Type~IIP SNe has found that their ejecta generally have aspherical structures especially in the inner core, occasionally extending also to the hydrogen envelope \citep[e.g.,][for a review]{Leonard2001, Leonard2006, Chornock2010, Nagao2019, Wang2008}. Recently, \citet[][]{Nagao2019} reported an unprecedented, highly-extended aspherical explosion in the Type~IIP SN 2017gmr indicated by an early rise of polarization, i.e., asymmetries are present not only in the helium core but also in a substantial portion of the hydrogen envelope. This implies a very aspherical explosion \citep[e.g., a jet-driven explosion;][]{Khokhlov1999}. On the other hand, because of their low abundance, Type~IIL SNe have not been subject to extensive polarimetric observations especially during the faint tail phase. The only case of a Type~IIL SN that has been observed with polarimetry in the tail phase is SN~2013ej \citep[][]{Kumar2016}, where $R$-band imaging polarimetry has been conducted. However, it displayed some intermediate properties between those of Type~IIP and IIL SNe \citep[e.g.,][]{Bose2015, Huang2015, Dhungana2016, Valenti2016, Yuan2016}. SN~2013ej exhibited high continuum polarization at the photospheric ($\sim1$ per cent) and tail phase ($\sim2$ per cent) \citep[][]{Kumar2016, Mauerhan2017}, which implies an aspherical explosion like those of Type~IIP SNe. 

In this study, we present multi-epoch spectropolarimetry of two Type~IIL SNe, SN~2013ej and SN~2017ahn, obtained with the Very Large Telescope (VLT). In the  analysis, we restrict ourselves to the continuum polarization. SN~2013ej was discovered by the Lick Observatory Supernova Search on 25.45 July 2013 UT \citep[56498.45 MJD;][]{Kim2013}. It is located in Messier~74 (Galaxy Morphology: SA(s)c; from NED\footnote{NASA/IPAC Extragalactic Database}) at $z=0.002192\pm0.000003$ and receding with a velocity of $v_{\rm{gal}}=657\pm1$ km s$^{-1}$ \citep[][]{Lu1993}. The object was not detected on 23.54 July 2013 UT \citep[56496.54 MJD;][]{Shappee2013}. We adopt MJD 56496.9 as an estimated explosion date \citep[][]{Dhungana2016}. 
SN~2017ahn was discovered by the `Distance Less Than 40 Mpc' supernova search \citep[DLT40;][]{Tartaglia2018} on 8.29 February 2017 UT \citep[57792.29 MJD;][]{Tartaglia2017}. It is located in NGC~3318 (Galaxy Morphology: SAB(rs)b; from NED) at $z=0.009255\pm0.000021$ and receding with a velocity of $v_{\rm{gal}}=2775\pm6$ km s$^{-1}$ (from the HIPASS Catalog via NED). The object was classified as a Type~II SN within a day after discovery \citep[][]{Hosseinzadeh2017}. The last non-detection of the object was on 7.23 February 2017 UT (57791.23 MJD), i.e. about one day before the discovery \citep[][]{Tartaglia2017}. We adopt MJD 57791.76 as an estimated explosion date \citep[][]{Tartaglia2021}. 

In the following section, we provide details of the observations and data reduction processes. Section 3 describes the methods used to correct the observations for reddening and interstellar polarization. Section 4 presents and discusses the results, and the conclusions follow in Section 5.


\begin{table*}
      \caption[]{Log of the VLT/FORS2 observations of SN 2017ahn}
    $\displaystyle
         \begin{array}{cccccccc}
            \hline
            \noalign{\smallskip}
            \rm{Date} & \rm{MJD} & \rm{Phase}^{a} & \rm{Days\; from \;explosion}^{b} & \rm{Airmass} & \rm{Exp. \;time} & \rm{Pol.\;degree} & \rm{Pol. \;angle}\\
            (\rm{UT}) & (\rm{days}) & (\rm{days}) & (\rm{days}) & (\rm{average}) & (\rm{s}) & (\rm{per\;cent}) & (\rm{degrees}) \\
            \noalign{\smallskip}
            \hline\hline
            \noalign{\smallskip}      
            2017-03-24.75 & 57836.75 & -16.63 & +44.99 & 1.1 & 4 \times 2076 & 0.09 \pm 0.07 & 27.9 \pm 20.3\\
            \noalign{\smallskip} \hline \noalign{\smallskip}
            2017-03-31.38 & 57843.38 & -10.00 & +51.62 & 1.2 & 4 \times 2076 & 0.11 \pm 0.09 & 50.6 \pm 28.2\\
            \noalign{\smallskip} \hline \noalign{\smallskip}
            2017-04-24.06 & 57867.06 & +13.68 & +75.30 & 1.1 & 4 \times 2076 & 0.16 \pm 0.14 & 43.4 \pm 34.5\\
            \noalign{\smallskip} \hline \noalign{\smallskip}
            2017-04-28.73 & 57871.73 & +18.35 & +79.97 & 1.2 & 4 \times 2076 & 0.18 \pm 0.15 & 51.7 \pm 35.4\\
            \noalign{\smallskip}
            \hline
         \end{array}
         $
         \begin{minipage}{.88\hsize}
        \smallskip
        Notes. ${}^{a}$Relative to $t_{0}=57853.38$ (MJD), which is the time of the end of the photospheric phase. ${}^{b}$Relative to $t=57791.76$ (MJD), which is the estimated time of the explosion. The error bars of the polarization degrees and angles represent the photon shot noise per bin. The error estimates do not include systematic effects such as the determination of the ISP.
        \end{minipage}
        \label{table1}
\end{table*}

\begin{table*}
      \caption[]{Log of the VLT/FORS2 observations of SN 2013ej}
    $\displaystyle
         \begin{array}{cccccccc}
            \hline
            \noalign{\smallskip}
            \rm{Date} & \rm{MJD} & \rm{Phase}^{a} & \rm{Days\; from \;explosion}^{b} & \rm{Airmass} & \rm{Exp. \;time} & \rm{Pol.\;degree} & \rm{Pol. \;angle}\\
            (\rm{UT}) & (\rm{days}) & (\rm{days}) & (\rm{days}) & (\rm{average}) & (\rm{s}) & (\rm{per\;cent}) & (\rm{degrees}) \\
            \noalign{\smallskip}
            \hline\hline
            \noalign{\smallskip}      
            2013-08-01.38 & 56505.38 & -91.99 & +8.48   & 1.4 & 4 \times 720  & 0.48 (0.00) \pm 0.04 & 72.3 (-) \pm 3.3\\
            \noalign{\smallskip} \hline \noalign{\smallskip}
            2013-08-27.28 & 56531.28 & -66.09 & +34.38  & 1.5 & 4 \times 900  & 0.81 (0.37) \pm 0.07 & 80.1 (90.0) \pm 1.2\\
            \noalign{\smallskip} \hline \noalign{\smallskip}
            2013-09-17.21 & 56552.21 & -45.16 & +55.31  & 1.5 & 4 \times 920  & 1.07 (0.68) \pm 0.19 & 85.6 (94.9) \pm 1.5\\
            \noalign{\smallskip} \hline \noalign{\smallskip}
            2013-09-29.24 & 56564.24 & -33.13 & +67.34  & 1.4 & 4 \times 1480 & 1.26 (0.91) \pm 0.21 & 89.3 (98.0) \pm 1.8\\
            \noalign{\smallskip} \hline \noalign{\smallskip}
            2013-10-29.23 & 56594.23 & -3.14  & +97.33  & 1.4 & 4 \times 1800 & 1.19 \pm 0.21 & 94.2 \pm 3.7\\
            \noalign{\smallskip} \hline \noalign{\smallskip}
            2013-12-04.56 & 56630.56 & +33.19 & +133.66 & 1.4 & 4 \times 3650 & 0.52 \pm 0.13 & 99.5 \pm 11.6\\
            \noalign{\smallskip} \hline \noalign{\smallskip}
            2014-01-09.38 & 56666.38 & +69.01 & +169.48 & 1.5 & 4 \times 1800 & 0.55 \pm 0.24 & 119.5 \pm 15.9\\
            \noalign{\smallskip}
            \hline
         \end{array}
         $
         \begin{minipage}{.88\hsize}
        \smallskip
        Notes. ${}^{a}$Relative to $t_{0}=56597.37$ (MJD), which is the time of the end of the photospheric phase. ${}^{b}$Relative to $t=56496.90$ (MJD), which is the estimated time of the explosion. The values in brackets for the polarization degree and angle refer to the continuum polarization after subtraction of the interaction component, i.e., they concern the aspherical explosion component. The error bars of the polarization degrees and angles represent the photon shot noise per bin. The error estimates do not include systematic effects such as the determination of the ISP.
        \end{minipage}
        \label{table2}
\end{table*}

\section{Observations and data reduction}

We obtained spectropolarimetry of the Type~IIL SN~2017ahn with the FOcal Reducer/low-dispersion Spectrograph 2 \citep[hereafter FORS2;][]{Appenzeller1998} mounted at the Cassegrain focus of VLT UT1. The observations were conducted from $\sim45$ to $\sim80$ days after the discovery and cover the photospheric phase and the transition phase to the tail phase. The observing log is presented in Table~\ref{table1}, where the phase is measured from the end of the photospheric phase, $t_{0}$. This was calculated using the $V$-band LC taken from \citet[][]{Tartaglia2021}, as in \citet[][]{Valenti2016}, i.e., as the midpoint of the luminosity drop from the photospheric to the tail phase: $t_{0}=57853.38$ (MJD). It is noted that this definition of $t_{0}$ and the phases derived from it differ from those presented in \citet[][]{Nagao2019}. We also analyzed the spectropolarimetric data of SN~2013ej taken with FORS2/VLT, which we obtained from the ESO Science Archive Facility. The observing log is presented in Table~\ref{table2}, where the phase is counted from the end of the photospheric phase, $t_{0}=56597.37$ (MJD). This was determined as for SN~2017ahn, using the $V$-band LC from the Lick Observatory Supernova Search program \citep[][]{deJaeger2019}. The data of the $V$-band LC of SN~2013ej were retrieved from the Berkeley SuperNova DataBase \citep[SNDB;][]{Silverman2012}.

The details of the spectropolarimetry and its analysis are the same as in \citet[][]{Nagao2019}: in FORS2, the spectrum generated by a grism is divided by a Wollaston prism into two beams with orthogonal polarization directions, i.e., the ordinary (o) and extraordinary (e) beams. The beam splitter is coupled to a half-wave retarder plate (HWP), which enables to measure the mean electric field intensity along different angles on the plane of the sky. For our observations, the optimal angle set $0^{\circ}$, $22.5^{\circ}$, $45^{\circ}$ and $67.5^{\circ}$ was adopted \citep[see][for more details]{Patat2006}. The HWP angle is measured as an angle between the fast axis of the HWP and the acceptance axis of the ordinary beam in the Wollaston prism, which is aligned to the north-south direction. As the dispersive element, we used the low-resolution G300V grism, which gave a spectral range of $3800-9200$ Angstrom, a dispersion of $\sim3.3$ Angstrom pixel$^{-1}$, and a resolution of $\sim 10$ Angstrom ($FWHM$) at $\sim 5800$ Angstrom. The data were reduced with IRAF\footnote{IRAF is distributed by the National Optical Astronomy Observatory, which is operated by the Association of Universities for Research in Astronomy (AURA) under a cooperative agreement with the National Science Foundation.} using the normal methods as described, e.g., in \citet[][]{Patat2006}. We extracted the ordinary and extraordinary beams of the spectropolarimetric data with a fixed aperture size of $10$ pixels, and then rebinned the extracted spectra to 50 Angstrom bins in order to improve the signal-to-noise ratio. Using the table data in \citet[][]{Jehin2005}, we corrected the HWP zero-point angle chromatism. We corrected the wavelength scale to the rest frame based on the galaxy redshift after subtracting the interstellar polarization (ISP). Finally, using the standard method described in \citet[][]{Wang1997}, we subtracted the polarization bias in the polarimetric spectra.

\section{Reddening and interstellar polarization} 

We adopted the Milky Way (MW) reddening values of $E(B-V)=0.061$ mag and $E(B-V)=0.069$ mag towards SN~2013ej and SN~2017ahn, respectively \citep[][]{Schlafly2011}. The extinction within the host galaxy of SN~2017ahn is assumed as $E(B-V) \sim0.2$ mag derived by \citet[][]{Tartaglia2021} using NaI D lines based on \citet[][]{Poznanski2012}. For SN~2013ej, we ignored the reddening within the host galaxy, because no Na I D lines from the host were detected \citep[][]{Valenti2014}. The empirical relation by \citet[][]{Serkowski1975} indicates that the total ISP can reach $\sim0.7$ and $\sim3.0$ per cent for SN~2013ej and SN~2017ahn, respectively. These upper limits can be realized only when all dust grains contributing to the extinction are aligned in the same direction. In a more realistic approach, we estimate the ISP component from the polarization spectra using the same method as in \citet[][]{Nagao2019}, i.e., assuming complete depolarization at the emission peaks of strong P~Cygni lines.

\begin{figure*}
  \includegraphics[width=\columnwidth,trim=45 0 45 0,clip]{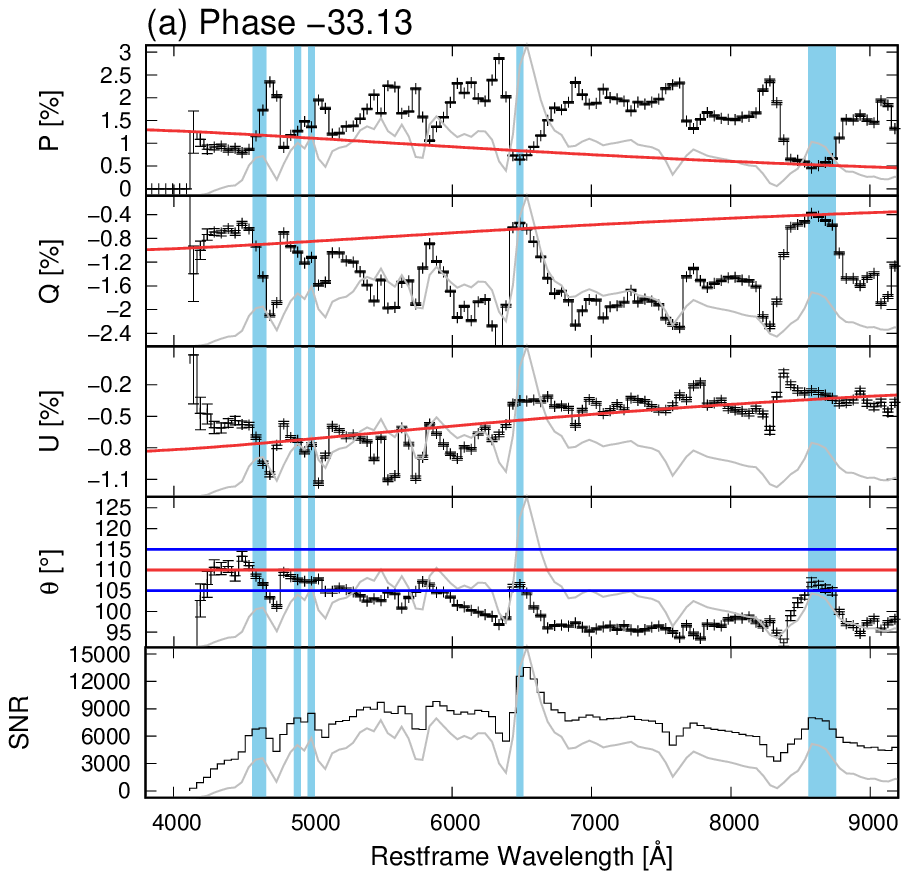}
  \includegraphics[width=\columnwidth,trim=45 0 45 0,clip]{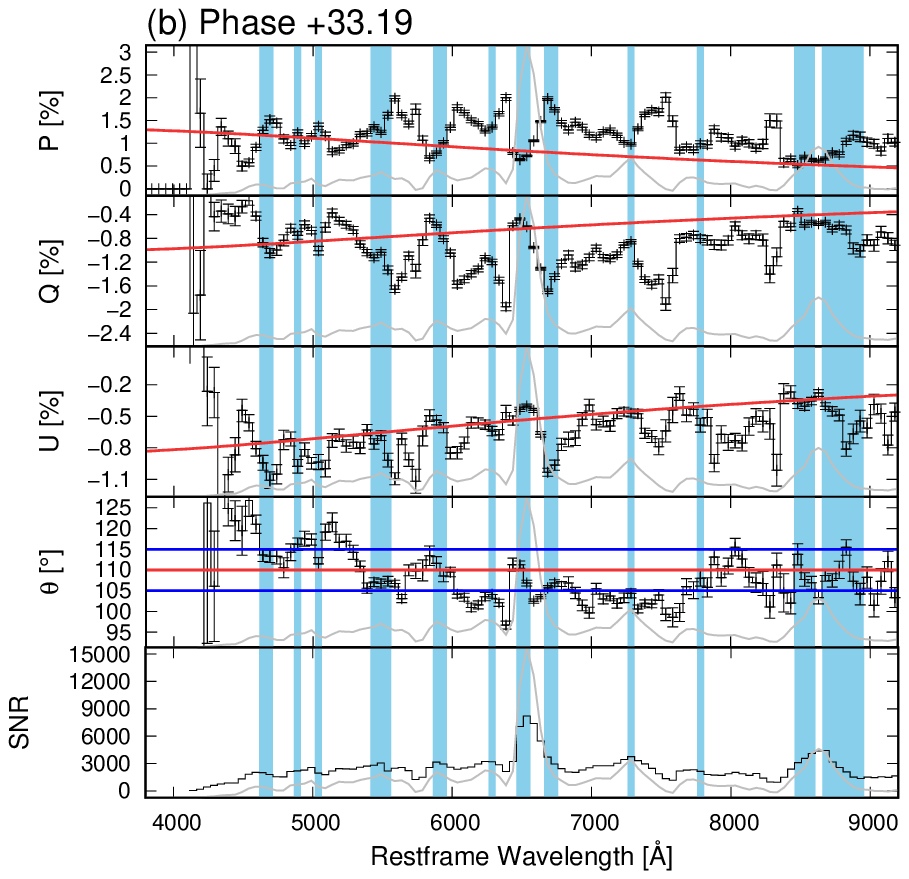}
    \caption{Polarization spectra of SN~2013ej before ISP subtraction at phases $-33.13$ (a) and $+33.19$ (b). From top to bottom, total polarization $P$, Stokes parameters $Q$ and $U$, polarization angle $\theta$, and signal-to-noise ratio (SNR) for SN~2013ej are plotted (black lines). The data were binned to $50$ Angstrom per point, and the SNR was evaluated as the photon shot noise per bin. The gray lines in the background of each plot are the unbinned flux spectra at the same epochs. The ISP is described by $P(\lambda) = P_{\rm{max}} \exp \left[ -K \ln^{2} \left( \lambda_{\rm{max}}/\lambda \right) \right]$, where $P_{\rm{max}}=1.32$, $\lambda_{\rm{max}}=3300$ Angstrom and $K=1.0$ (red lines). In the 4th panel (polarization angle), the red line and blue lines represent the assumed ISP angle ($\theta=110$ degrees) and the adopted maximum and minimum polarization angle ($\theta=105$ and $\theta=115$ degrees) for determining the ISP, respectively. The blue hatching identifies the adopted wavelength range for the ISP-dominated components.
    }
    \label{fig1}
\end{figure*}

\begin{figure*}
  \includegraphics[width=\columnwidth,trim=45 0 45 0,clip]{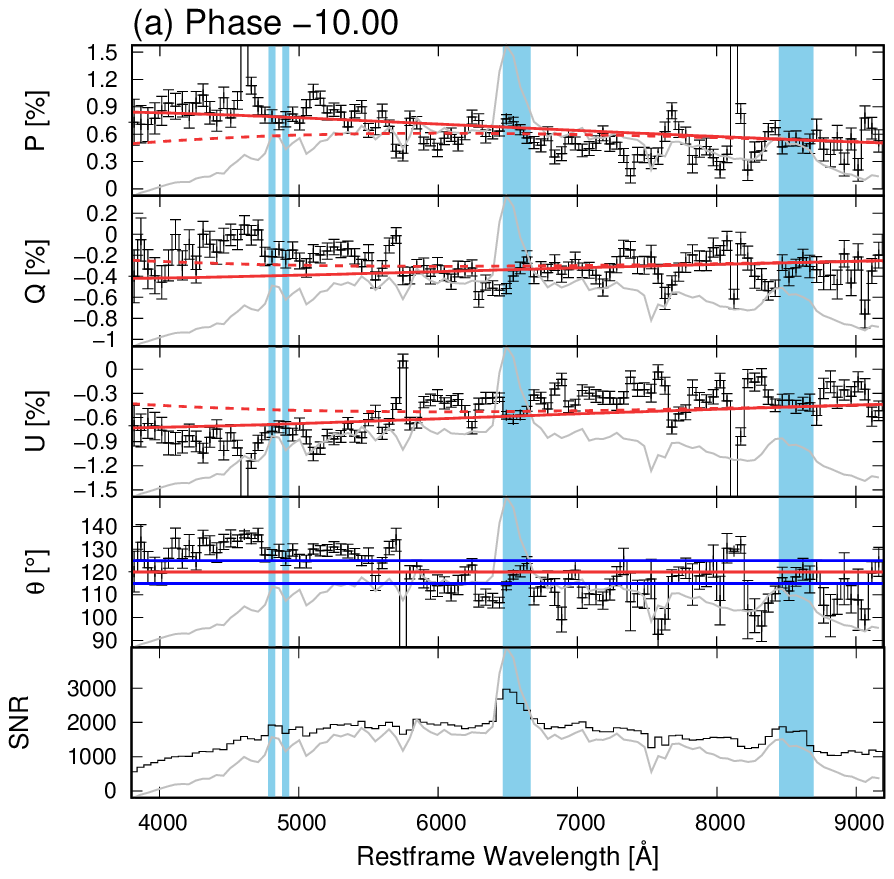}
  \includegraphics[width=\columnwidth,trim=45 0 45 0,clip]{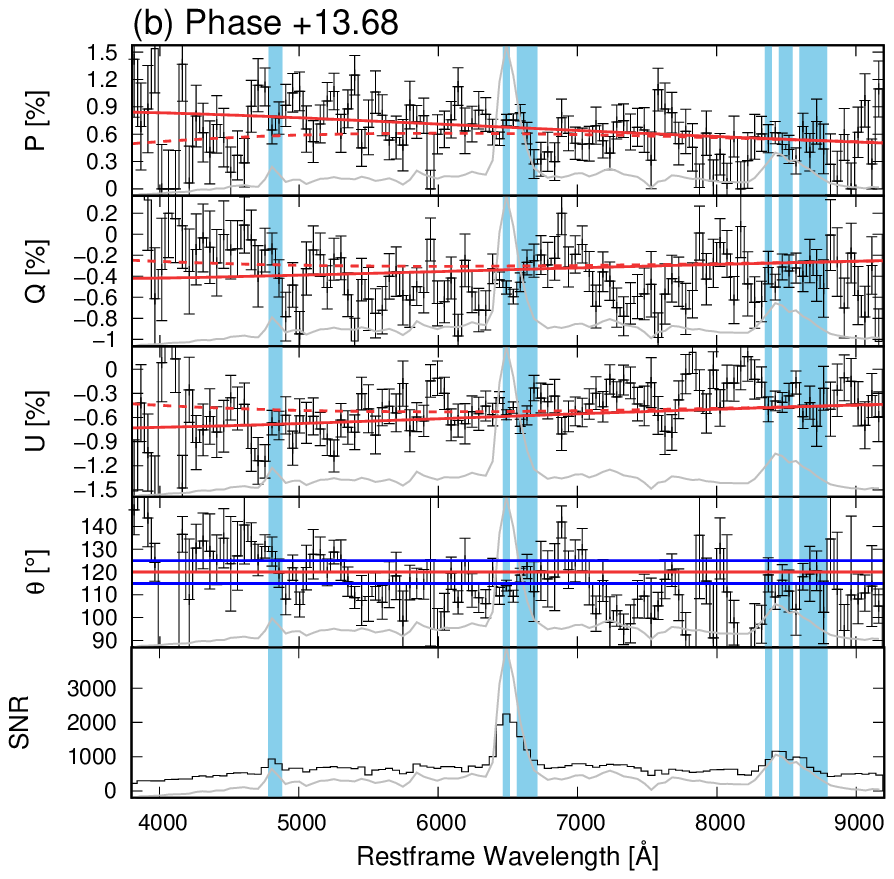}
    \caption{Same as Fig.~\ref{fig1}, but for SN~2017ahn at phases $-10.00$ (a) and $+13.68$ (b). The ISP is described by $P(\lambda) = P_{\rm{max}} \exp \left[ -K \ln^{2} \left( \lambda_{\rm{max}}/\lambda \right) \right]$. The solid red lines show the case with $P_{\rm{max}}=2.05$, $\lambda_{\rm{max}}=200$ Angstrom and $K=0.1$, while the dotted red lines represent the case with $P_{\rm{max}}=0.61$, $\lambda_{\rm{max}}=6000$ Angstrom, and $K=1.0$. Here, the assumed ISP angle is $\theta=120$ degrees. 
    }
    \label{fig2}
\end{figure*}

Figures \ref{fig1} and \ref{fig2} show polarization spectra of SN~2013ej (phases -33.13 and +33.19 days) and SN~2017ahn (phases -10.00 and +13.68 days), respectively, before ISP subtraction (polarization spectra for all epochs of our data are provided in Figs.~\ref{figa1} [SN~2013ej] and \ref{figa2} [SN~2017ahn] in Appendix~A). The error bars of the polarization degrees and angles represent the photon shot noise per bin. It is noted that the error estimates do not incorporate systematic effects such as the determination of the ISP. The emission peaks of the strong emission lines (e.g., H$\alpha$, Ca II triplet, etc.) show non-zero polarization that has a different angle from that in the other wavelength regions. This demonstrates the existence of another component with a different polarization angle from the intrinsic SN component. We assumed this non-intrinsic component as the ISP. The polarization angles around the emission peaks are around 110 and 120 degrees for SN~2013ej and SN~2017ahn, respectively, which we assume are the ISP angles. The ISP angle of SN~2017ahn seems to be aligned with the nearby spiral arm in the host galaxy, while that of SN~2013ej seems to be perpendicular.

The ISP wavelength dependency was obtained by fitting all the polarimetric data with the Serkowski function \citep[][]{Serkowski1975}: $P(\lambda) = P_{\rm{max}} \exp \left[ -K \ln^{2} \left( \lambda_{\rm{max}}/\lambda \right) \right]$, restricting the fit to wavelength windows where the error range of the polarization angle overlaps with the 5 degree range assumed for the ISP angle and where the flux is $> 1.1$ times higher than in the nearby continuum region (see the blue hatching in Figs.~\ref{fig1} and \ref{fig2}). The best-fit values for the fitting parameters are: $P_{\rm{max}}=1.32$ per cent, $\lambda_{\rm{max}}=3300$ Angstrom and $K=1.0$ (SN~2013ej) and $P_{\rm{max}}=2.05$ per cent, $\lambda_{\rm{max}}=200$ Angstrom 
and $K=0.1$ (SN~2017ahn; see red lines in Figs.~\ref{fig1} and \ref{fig2}). With these numbers, $P_{\rm{ISP}}$, $Q_{\rm{ISP}}$ and $U_{\rm{ISP}}$ at $\lambda=8000$ Angstrom are $0.60$, $-0.46$ and $-0.39$ per cent for SN~2013ej, and $0.53$, $-0.26$ and $-0.46$ per cent for SN~2017ahn, respectively. It is noted that the derived values of $\lambda_{\rm{max}}$ and $K$ for SN~2017ahn are extreme compared with the extinction for MW stars \citep[$\lambda_{\rm{max}}\sim6000$ and $K \sim1.0$; e.g.,][]{Whittet1992}, but they are just best-fit values. In fact, the range of the 90 per cent confidence level also includes more familiar values (see Nagao et al. 2021 [in prep.] for details on the ISP components). For example, in the above fitting, we obtain $P_{\rm{max}}=0.61$ per cent with fixed values of $\lambda_{\rm{max}}=6000$ Angstrom and $K = 1.0$. These values do not change the derived ISP by more than the 1-$\sigma$ error bars of our observations at the redder wavelengths of interest (see the red dotted lines in Fig.~\ref{fig2}). On the other hand, the ISP estimation in the bluer region is less robust because there are no prominent lines. In this paper, we use the best-fit values for the ISP subtraction. Stokes Q-U diagrams for SN~2013ej and SN~2017ahn are presented in Appendix~A.

As the ISP in SN~2013ej, \citet[][]{Mauerhan2017} adopted that derived for SN~2002ap \citep[wavelength-independent  values of $P_{\rm{ISP}}=0.51$ per cent and $\theta_{\rm{ISP}}=125$ degrees;][]{Leonard2002}, which happened in the same host galaxy.  They also assumed that the host ISP for SN~2013ej and SN~2002ap is negligible. Their values of $P_{\rm{ISP}}$ and $\theta_{\rm{ISP}}$ are somewhat different from our values derived directly from the spectropolarimetry of SN~2013ej ($P_{\rm{ISP}} = 0.60$ per cent and $\theta_{\rm{ISP}}=110$ degrees at $\lambda=8000$\,\AA). From spectropolarimetric observations of SN~2002ap, \citet[][]{Wang2003} pointed out that the host ISP towards SN~2002ap dominates over the MW ISP. On the other hand, the major extinction towards SN~2013ej should originate from MW dust, as indicated by the Na line (see above). Thus, contrary to SN~2002ap, the MW ISP might be stronger than the host ISP towards SN~2013ej.  Even if the main origin of the ISP towards SN~2013ej is the relatively little dust in the host galaxy, owing to the $\sim 8$ kpc physical distance between SN~2013ej and SN~2002ap the line-of-sight dust in the host may have different properties. Therefore, the ISP adopted by \citet[][]{Mauerhan2017} might not be appropriate for SN~2013ej. In fact, after ISP subtraction, the emission peaks in the polarimetric spectra of \citet[][]{Mauerhan2017} do not reach the zero level (see Fig.\,2 in \citet[][]{Mauerhan2017}). On the other hand, \citet[][]{Kumar2016}, for SN~2013ej, used the MW ISP derived from the broadband imaging polarimetry of foreground stars in the vicinity of the SN position ($P_{\rm{ISP}} = 0.64$ per cent and $\theta_{\rm{ISP}}=108.1$ degree in the $R$ band). These values are similar to our estimates.

\section{Results and discussion}

\subsection{SN~2013ej}

\begin{figure*}
  \includegraphics[width=1.8\columnwidth]{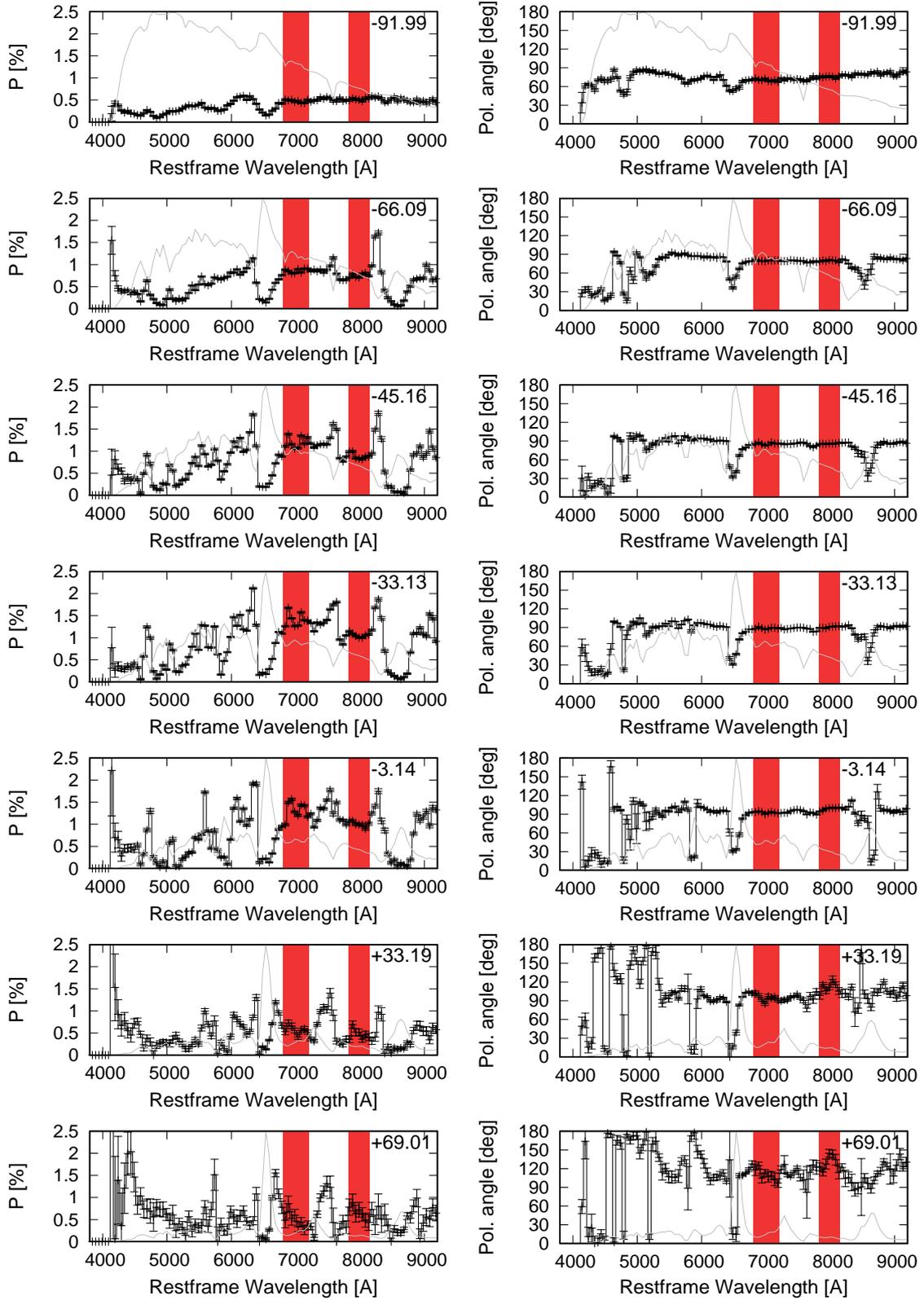}
    \caption{Polarization degree (left) and angle (right) of SN~2013ej after ISP subtraction at different epochs (increasing from top to bottom as labeled). The gray lines in the background of each plot are the unbinned flux spectra at the same epochs. The red hatching shows the adopted wavelength range for the continuum polarization estimate.  
    }
    \label{fig3}
\end{figure*}

\begin{figure*}
\includegraphics[width=\columnwidth]{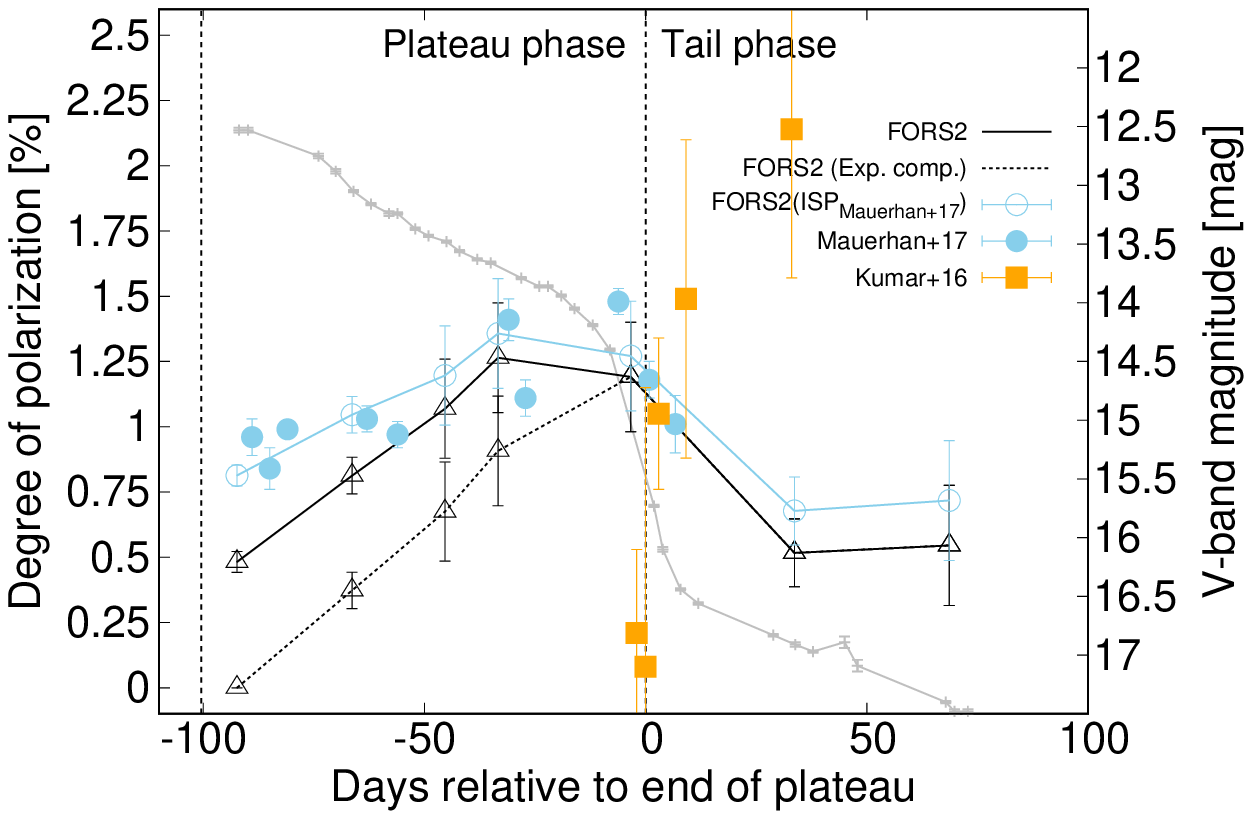}
\includegraphics[width=\columnwidth]{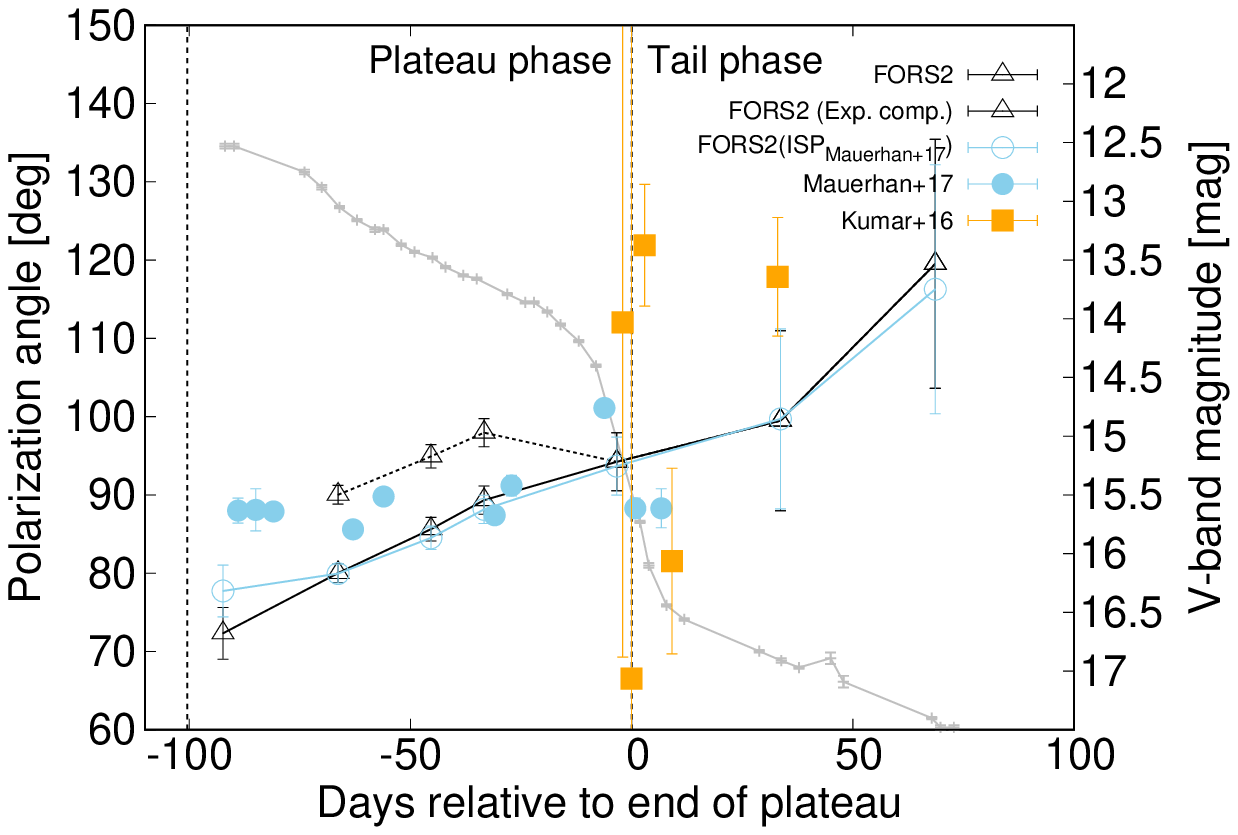}
    \caption{Time evolution of the continuum polarization of SN~2013ej, compared with previous studies. The left and right panels show the polarization degree and angle, respectively. The black open triangles connected by a solid line trace the derived total intrinsic continuum polarization (the sum of 
    the interaction and explosion components). The black open triangles connected by a dotted line indicate the explosion component derived by subtracting the interaction component from the total. The cyan filled circles are the continuum polarization estimated by \citet[][]{Mauerhan2017}, while the yellow filled squares are the $R$-band polarization obtained by \citet[][]{Kumar2016}. The cyan open circles are FORS2 spectropolarimetric data (using the same ISP and wavelength range of the continuum polarization as in \citet[][]{Mauerhan2017}). The gray crosses connected by a line trace the $V$-band LC of SN~2013ej taken from \citet[][]{deJaeger2019}.
    }
    \label{fig4}
\end{figure*}

\begin{figure}
  \includegraphics[width=\columnwidth]{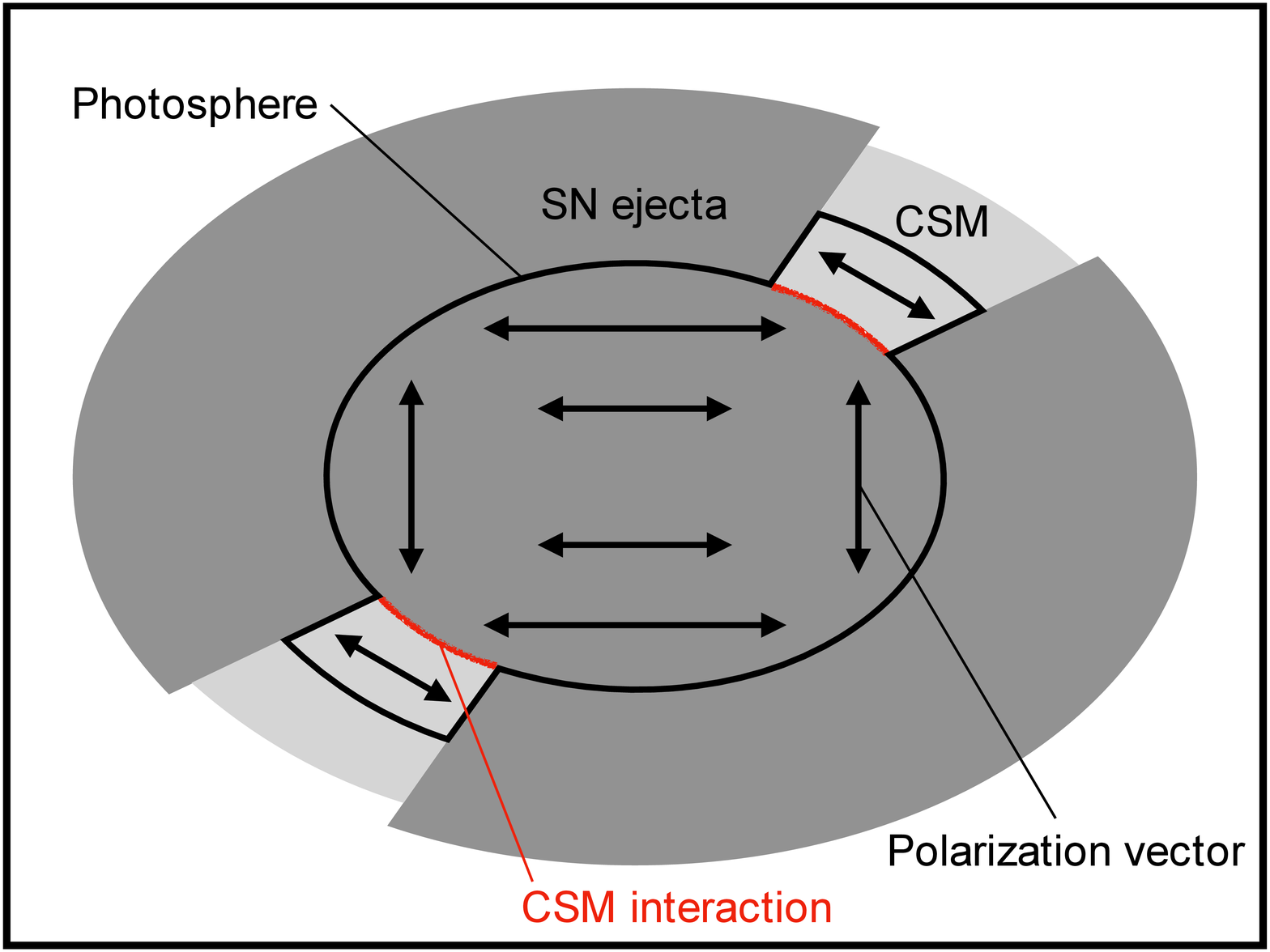}
    \caption{Schematic presentation of the continuum polarization in SN~2013ej (observer's view), which can be broken down into a component produced by the aspherical SN ejecta (the explosion component) and another one arising from an asymmetric interaction zone (the interaction component).
    }
    \label{fig5}
\end{figure}

Figure~\ref{fig3} presents the polarization degree and angle of SN~2013ej after ISP subtraction. Especially in the redder wavelength region (e.g., $\gtrsim 6000$ Angstrom), SN~2013ej shows continuum polarization with a constant polarization angle at all epochs whereas the continuum polarization especially in the bluer wavelength region (e.g., $\lesssim 6000$ Angstrom) is heavily contaminated by line polarization. For the determination of the continuum polarization, we adopted the wavelength ranges between 6800 and 7200 Angstrom as well as between 7820 and 8140 Angstrom, following \citet[][see the red hatched region in Fig.~\ref{fig3}]{Chornock2010}. Mean values of the continuum polarization in these wavelength regions are presented in Table~\ref{table2} and Fig.~\ref{fig4}. As can be seen in Fig.~3, there is intrinsic scatter across the adopted continuum regions that exceeds the photon shot noise. This indicates the presence of other components such as line polarization so that photon shot noise is not a sufficient basis for error estimates. Therefore, for the calculation of the errors of the continuum polarization, we adopt the square root of the squared sums of the photon noise and of the standard deviation of the 50-Angstrom-binned data in the continuum regions. The results are included in Table~\ref{table2}.  

We note that our values of the continuum polarization are slightly different from the previous results by \citet[][]{Mauerhan2017} and \citet[][]{Kumar2016}. \citet[][]{Mauerhan2017} presented spectropolarimetry of the photospheric phase taken with the Shane 3m reflector and the Kast spectrograph at Lick Observatory and reported a higher polarization degree and different polarization angles (see the cyan filled circles in Fig.~\ref{fig4}). To check this discrepancy, we calculated the continuum polarization from our data with the same ISP and continuum region (the wavelength range between 7800 Angstrom and 8150 Angstrom) as used by \citet[][]{Mauerhan2017}. Since their adopted continuum region is too narrow to evaluate the scatter of the 50-Angstrom binned data, we adopted the same errors as in Table~2 for the calculated polarization degrees and angles. The results are closer to their values, although the derived polarization angles are still different from theirs (see the cyan open circles in Fig.~\ref{fig4}). \citet{Mauerhan2017}'s own error estimates only include the photon noise. However, as for our observations, this is almost certainly an underestimate. In summary, we conclude that the discrepancy can be attributed to the different ISP assumptions and to the different error estimations. But the different data quality imprinted by the different light-collecting areas of the telescopes may also be a relevant factor.

\citet[][]{Kumar2016} presented $R$-band imaging polarimetry in the tail phase. Their derived values for the continuum polarization are substantially different from ours and also from those in \citet[][see Fig.~\ref{fig4}]{Mauerhan2017}, and their observations have large uncertainties. Especially for the first and last epochs, the measurements by \citet{Kumar2016} are inconsistent with ours. Since they use broad-band imaging polarimetry, the impact of non-photon-shot noise should be larger. In fact, their published polarization angles scatter by more than the error bars, which indicates that the uncertainties might be a bit larger than reported. In conclusion, the VLT observations appear to be the most reliable. Therefore, we base our discussion of the continuum polarization of SN~2013ej and its evolution on these data only.

During the early photospheric phase, SN~2013ej exhibited a high continuum polarization level that was never before observed in other Type~IIP/L SNe. The polarization increased through the end of the photospheric phase, and the polarization angle also changed. When the SN entered the tail phase, the polarization degree decreased. There are two possible sources of continuum polarization in Type~II SNe: electron scattering in an aspherical photosphere (the aspherical-photosphere scenario) and dust scattering in asymmetrically distributed circumstellar matter \citep[the dust scattering scenario; see also the discussion in][and references therein]{Nagao2019}. \citet[][]{Nagao2018} have demonstrated that the wavelength dependence of the continuum polarization in the dust scattering scenario is always different from that in the aspherical photosphere scenario. The polarization expected from an aspherical photosphere is wavelength-independent, because electron scattering does not depend on wavelength. On the other hand, the polarization by dust scattering always carries some wavelength dependence. Roughly speaking, the polarization is higher in the shorter wavelength region up to the wavelength where the optical depth of the CSM is $\sim2.0$.  Below this wavelength, the polarization decreases towards shorter wavelengths due to multiple scattering \citep[][]{Nagao2018}. In most cases of the CSM expected from red supergiants \citep[e.g., with a mass-loss rate of $\lesssim 10^{-5}$ $\rm{M}_{\odot}\rm{yr}^{-1}$;][]{Smith2014b}, the polarization, in the optical bands, is higher in the shorter wavelength region. Since the polarization spectra for SN~2013ej do not show such wavelength dependence in the wavelength region without lines at any epoch (Fig.~\ref{fig3}), we conclude that the detected polarization is produced by an aspherical photosphere.

The comparatively high polarization in the very early phase indicates that the ejecta of SN~2013ej are quite aspherical in the outermost layer. The polarization degree at the first epoch ($\sim0.5$ per cent) implies an aspherical structure with an axis ratio of 1.1:1 in the electron-scattering atmosphere model by \citet[][]{Hoflich1991}. We note that this axis ratio is the minimum value assuming that the SN is viewed from the equatorial plane. If the line of sight is closer to the polar axis of the structure, larger asphericities are required. 
As noted above, such a highly aspherical structure in the outermost layer of the ejecta has never been observed in Type~IIP/L SNe, 
although some large aspherical structures were suggested for some Type~IIn SNe to have resulted from interaction with an aspherical CSM \citep[e.g., SN~2009ip;][]{Mauerhan2014,Reilly2017}. As \citet[][]{Mauerhan2017} pointed out, explanation of this extreme aspherical structure requires an interaction with highly aspherical CSM or a very significantly aspherical explosion (e.g., a jet-driven explosion), or both. 

Another important fact is that the polarization angle evolves with time, which, in the absence of any not physically motivated structural twisting, implies the presence of multiple polarization components characterized by different angles. Since it is difficult to explain this multi-directional structure merely with a very significantly aspherical explosion (e.g., a jet-driven explosion), we interpret the polarization of SN~2013ej as originating from the combination of an aspherical CSM interaction (the interaction component) and an aspherical explosion (the explosion component). In this respect we note that there is other observational evidence supporting an aspherical CSM interaction, e.g., late-time optical imaging and spectroscopy and X-ray observations \citep[see][and references therein]{Mauerhan2017}. Moreover, strong CSM interaction in the photospheric phase has also been proposed as a general explanation of the fast declining LC shapes in Type~IIL SNe (see \S 1). This CSM should have a small enough radial extent to be immediately swept up because normally we do not observe obvious interaction features (e.g., a narrow H$\alpha$ line, possibly with a Lorentzian wing) that are characteristic of optical spectra of Type IIn SNe (see references in \S 1).

In the following we assume that the polarization degree and angle at the first epoch ($P=0.48$ per cent and $\theta=72.3$ degrees) arise solely from an aspherical CSM interaction and that this interaction component survives until the beginning of the luminosity drop into the tail phase (from the first to the fourth epoch in our dataset). On this basis, we derive the explosion component of the polarization by subtracting the interaction component from the total intrinsic continuum polarization. The resulting explosion component shows a constant polarization angle ($\sim95$ degrees) that is different from that of the interaction component ($\theta=72.3$ degrees) except during the last epoch (the black open triangles connected by a dotted line in Fig.~\ref{fig4}). We note that the continuum polarization in the last epoch is obviously contaminated by line polarization, where the polarization degree and angle show a marked dispersion even within the continuum region (see Fig.~\ref{fig3}). This is because, in the later phases, the photosphere in the SN ejecta is becoming less defined and thus line polarization dominates over the continuum polarization.

The polarization level of the derived explosion component shows an increase during the photospheric phase, a peak around the beginning of the tail phase and a decline in the tail phase: $P_{\rm{max}}\sim1.2$ per cent and $\theta \sim95$ degrees (an average value from the second to the sixth epoch). The early rise of the of the explosion component of the polarization degree at constant polarization angle implies an extended aspherical explosion structure. This behavior is similar to that observed in Type~IIP SNe, especially in SN~2017gmr. In summary, the observed continuum polarization is interpreted as the combination of an interaction component from an aspherical CSM interaction ($P=0.48 \%$ and $\theta=72.3$ degrees) plus an explosion component from an aspherical explosion ($P_{\rm{max}}\sim1.2 \%$ and $\theta\sim95$ degrees; see Fig.~\ref{fig5}).

Finally, we also note that the LC of SN~2013ej differs from that of SN~2017ahn in that SN~2013ej exhibits a broader shoulder just after the peak (i.e., the peak of SN~2013ej appears flatter; see the gray lines in Figs.~\ref{fig4} and \ref{fig7}). This might be a another symptom of the CSM interaction.

\subsection{SN~2017ahn}

\begin{figure*}
  \includegraphics[width=1.8\columnwidth]{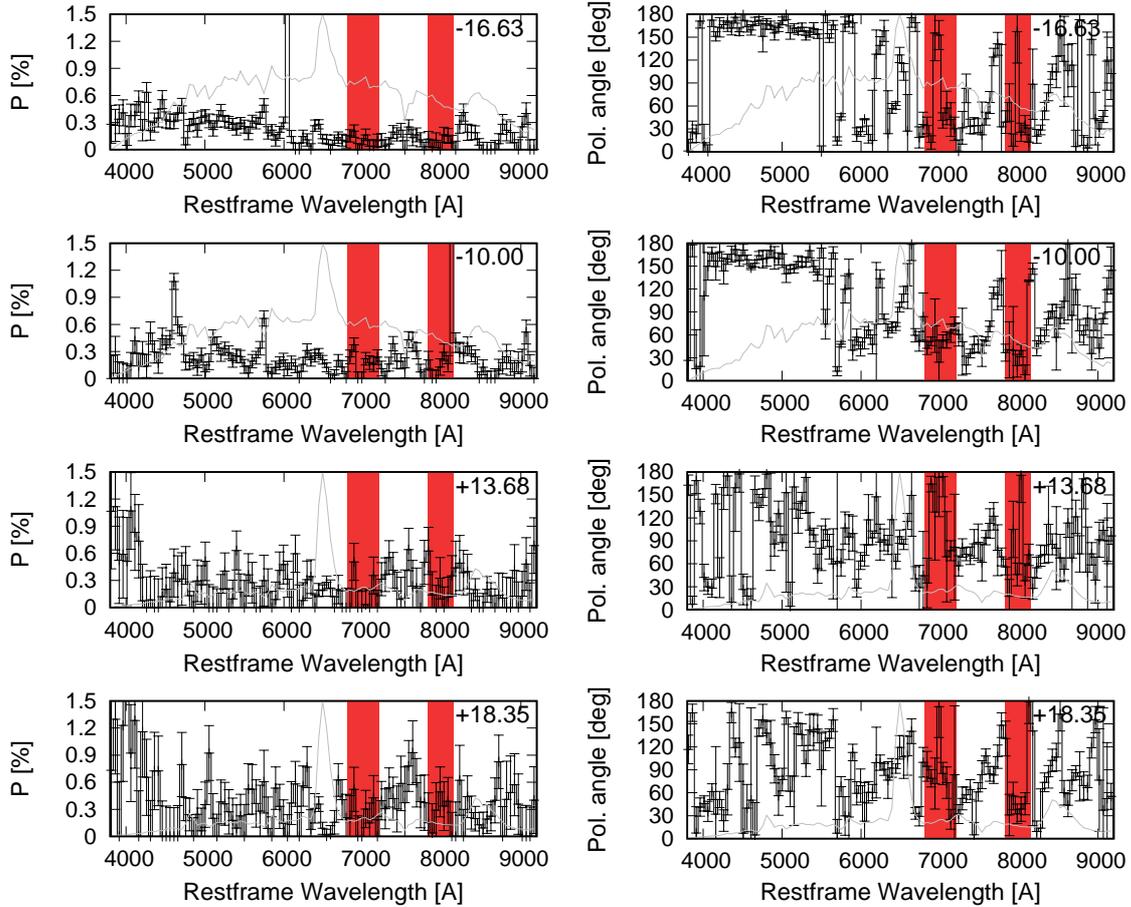}
    \caption{Polarization degree (left) and angle (right) of SN~2017ahn after ISP subtraction at different epochs (from top to bottom as labeled). The gray lines in the background of each plot are the unbinned flux spectra at the same epochs. The red hatching shows the wavelength range adopted for the continuum-polarization estimate.
    }
    \label{fig6}
\end{figure*}

\begin{figure*}
\includegraphics[width=\columnwidth]{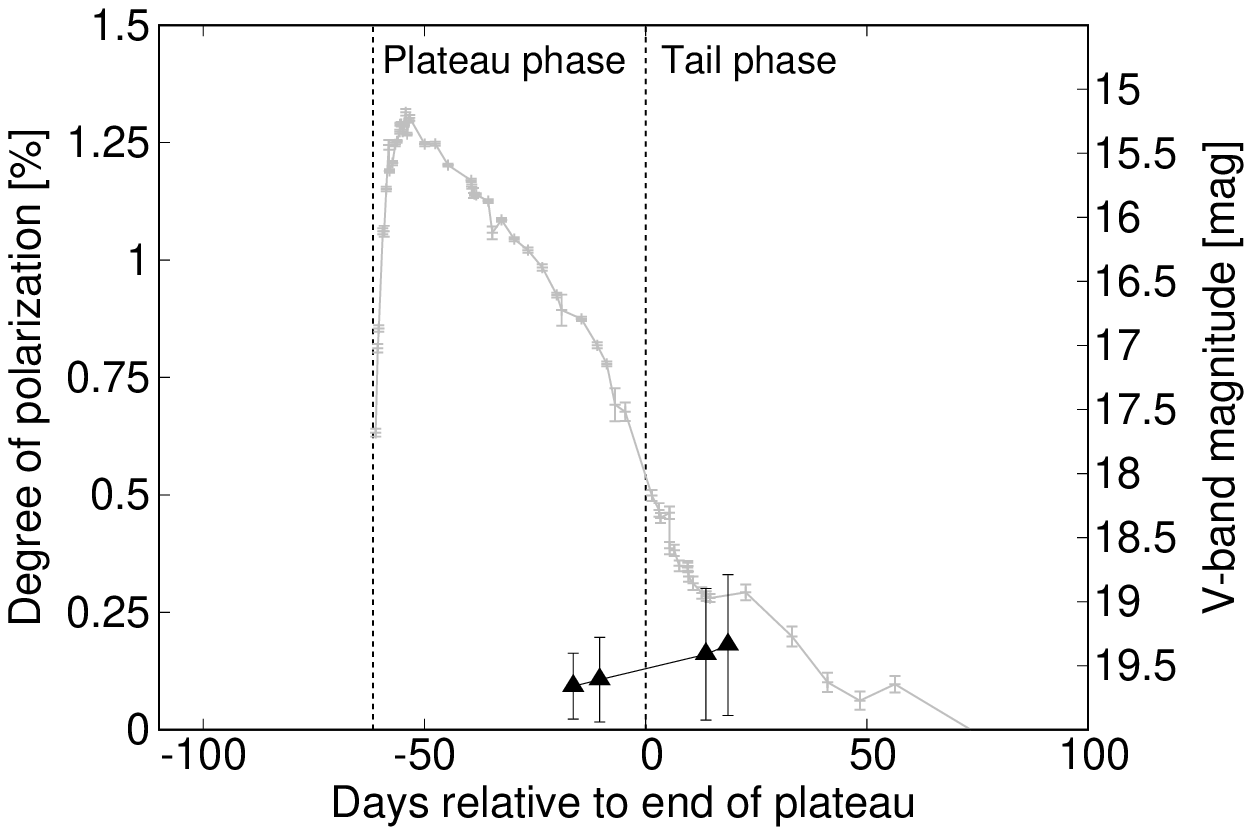}
\includegraphics[width=\columnwidth]{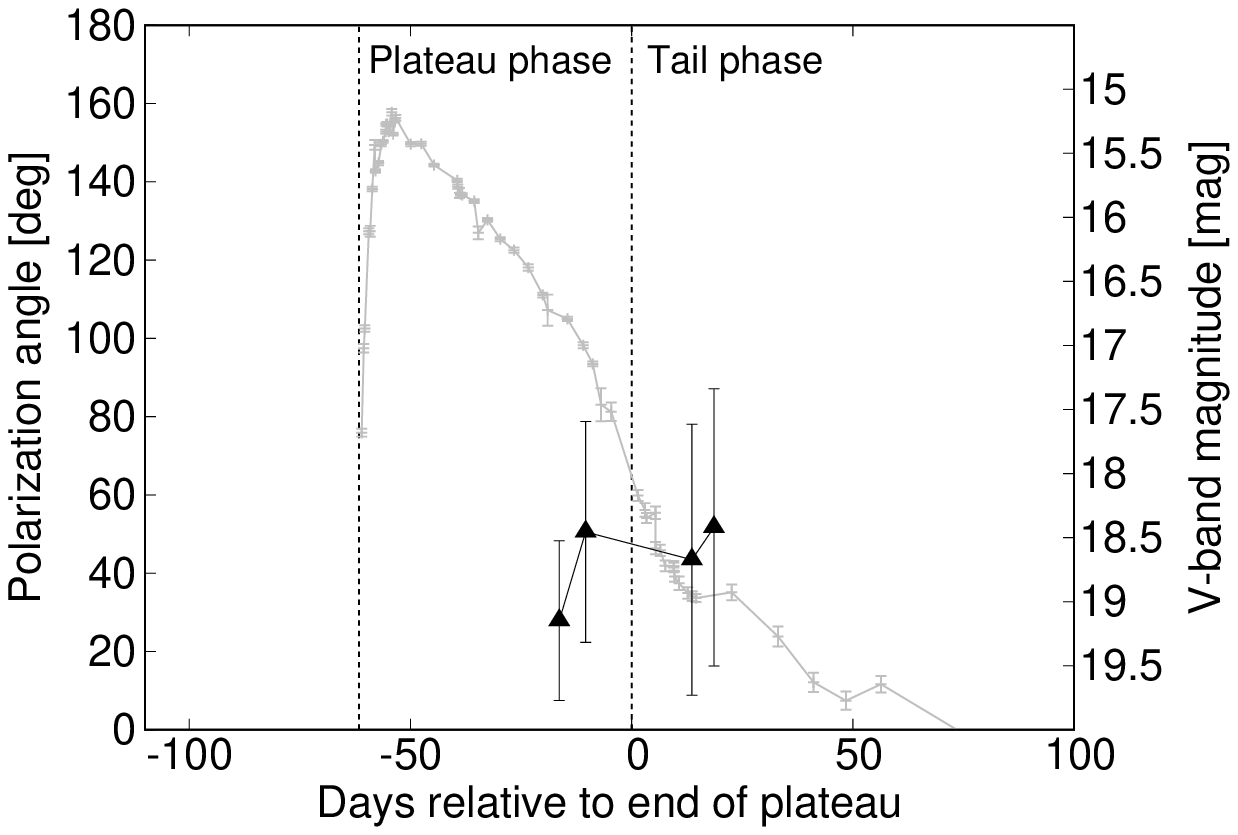}
    \caption{Time evolution of the continuum polarization of SN~2017ahn (red filled triangles). The gray crosses connected by a line show the $V$-band LC of SN~2017ahn taken from \citet[][]{Tartaglia2021}.
    }
    \label{fig7}
\end{figure*}

\begin{figure*}
\includegraphics[width=2.0\columnwidth]{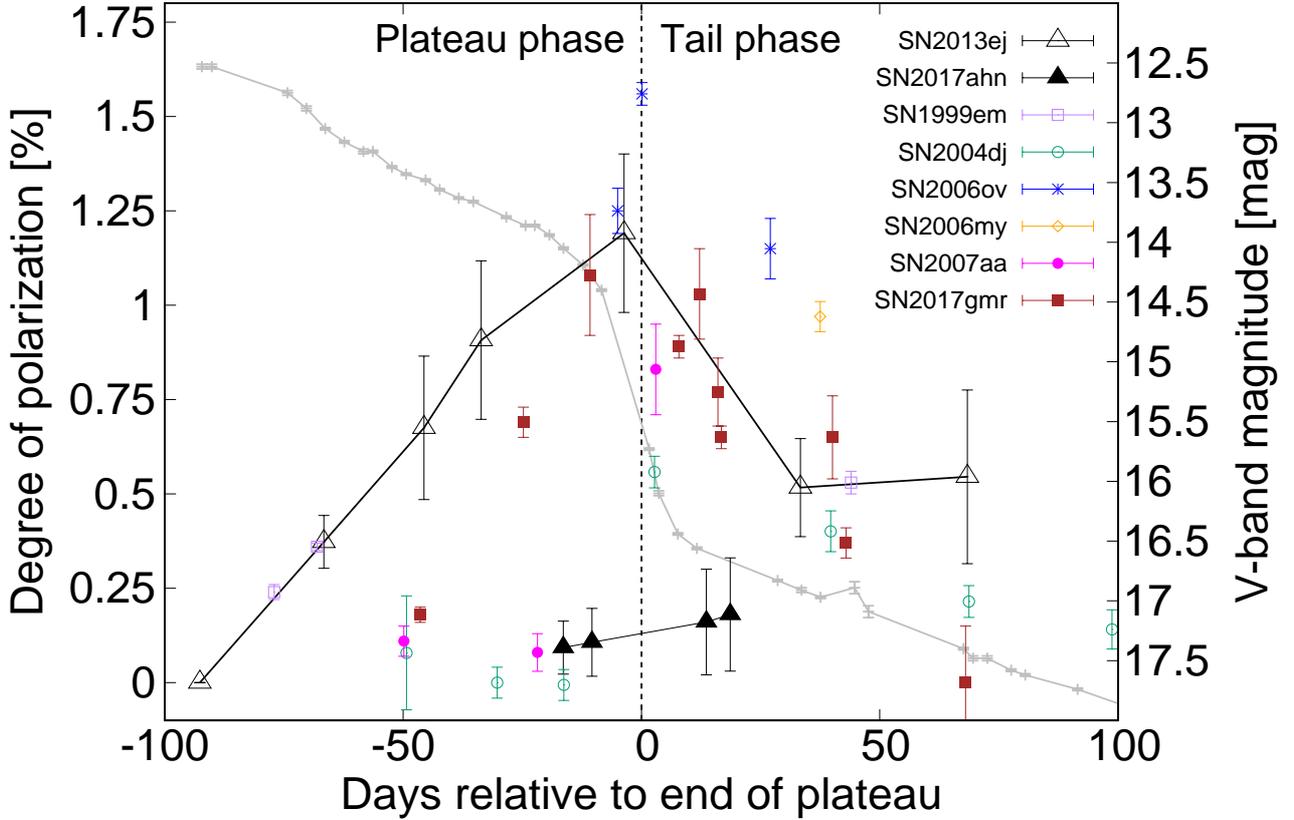}
    \caption{Continuum polarization of SN~2013ej and SN~2017ahn (black open and filled triangles, respectively), compared to other Type~IIP SNe. For SN~2013ej, only the explosion component is plotted. The gray crosses connected by a line show the $V$-band LC of SN~2013ej. The black vertical dashed line indicates the end of the photospheric phase: $t_{0}=56597.37$ (MJD) for SN~2013ej and $t_{0}=57853.38$ (MJD) for SN~2017ahn. The data for SNe 1999em, 2004dj, 2006ov, 2006my, 2007aa and 2017gmr (purple squares, green circles, blue stars, orange diamonds, magenta dots and brown squares, respectively) are taken from \citet[][]{Leonard2001, Leonard2006, Chornock2010, Nagao2019}. The adopted epochs for the end of the photospheric phase are: MJD 51597.4, 53288.3, 54094.5, 54084.0, 54227.3 and 58092.0 for SNe 1999em, 2004dj, 2006ov, 2006my, 2007aa and 2017gmr, respectively.
    }
    \label{fig8}
\end{figure*}

Figure~\ref{fig6} shows the ISP-subtracted polarization degree and angle of SN~2017ahn as a function of wavelength. The values of the continuum polarization are presented in Table~\ref{table1} and Fig.~\ref{fig7}. The errors were calculated in the same way as for SN~2013ej and are also included in Table~\ref{table1}. Since, contrary to SN~2013ej, these errors are photon-noise dominated, they should be upper limits.  This SN shows a very low polarization degree not only during the photospheric phase but also in the tail phase (Fig.~\ref{fig7}). In general, Type~IIP SNe show diversity in their polarization in the photospheric phase (see Fig.~\ref{fig8}), ranging from high \citep[e.g., SN~2017gmr and SN~2012aw;][]{Nagao2019, Dessart2021} to low levels \citep[e.g., SN~2004dj and SN~2007aa;][]{Leonard2006, Chornock2010}. However, they are typically characterised by high polarization at the onset of the tail phase, which implies that the inner cores are always, to some extent, aspherical \citep[see Fig.~\ref{fig4}; e.g.,][for a review]{Leonard2001, Leonard2006, Chornock2010, Nagao2019, Wang2008}. Therefore, the low polarization of SN~2017ahn, which is unprecedented in Type~IIP/L SNe, indicates spherical symmetry not only in the hydrogen envelope but also in the inner core: the axial ratio is only 1.06:1 according to the electron-scattering atmosphere model by \citet[][]{Hoflich1991}.

However, this low polarization could be explained in terms of viewing angle: SN~2017ahn could have a large asphericity similar to that of SN~2013ej and Type~IIP SNe if the line of sight were closer to the polar axis of the aspherical structure. If we assume that SN~2017ahn has the same aspherical structure as derived for SN~2017gmr \citep[an ellipsoid with an axial ratio of 1.2:1.0; see][]{Nagao2019}, then the viewing angle (measured from the polar axis) must be $\lesssim 30$ degree. In this respect we note that, in the case of a random orientation (i.e., if there is no relation between the LC shape and the viewing angle), the probability of a viewing angle $\lesssim 30$ degree is only $\sim15$ per cent. Since the assumed axial ratio is the minimum value for the aspherical structure in SN~2017gmr \citep[see][]{Nagao2019}, the realistic probability that SN~2017ahn is similarly aspherical as SN~2017gmr can be smaller.  

The low polarization in the photospheric phase suggests that, even if SN~2017ahn had a similar extensive CSM interaction as that of SN~2013ej, the interaction should have been relatively spherical. In fact, early flash-ionized spectra of SN~2017ahn suggest the existence of a confined dense CSM shell \citep[][]{Tartaglia2021}. However, it is unclear whether this interaction was strong enough to be the cause of the steep LC shape.

\subsection{The origin of the diversity of the LC shapes in Type~II SNe}

Unveiling the explosion geometry is important for understanding the origin of the diverse LC properties observed in Type~II SNe. As described in the Introduction, we consider two candidate causes for  rapidly-declining SN LCs:  (1) small hydrogen-envelope masses by enhanced mass loss inflicted by either a stellar wind or binary interaction or (2) interaction with CSM formed not too long before the explosion by either some instability of the progenitor or binary interaction. In both cases, the binary interaction scenario predicts similar progenitors for both Type~IIP and IIL SNe, while the stellar wind scenario (case 1) as well as the stellar instability scenario (case 2) require different progenitors.

SN~2013ej showed high polarization from soon after the explosion through the tail phase, with time-variable polarization angles indicating an aspherical explosion like in Type~IIP SNe and an interaction with an aspherical CSM. The similarity of the explosion geometries of SN~2013ej and Type~IIP SNe suggests that SN~2013ej had a progenitor similar to those of Type~IIP SNe. This implies that the steeper LC of SN~2013ej compared to Type~IIP SNe was caused mainly by either a small hydrogen-envelope mass or CSM interaction, in both cases as a result of binary interaction. The discrimination between these two possibilities in favor of CSM interaction is provided by the observed polarization. However, it still needs to be tested whether the CSM interaction that is invoked to explain the polarization can quantitatively account for the steepness of the LC of SN~2013ej.
In summary, SN~2013ej probably originated from a progenitor similar to those of Type~IIP SNe, but with a companion whose mass and/or separation was different from those in typical SNe IIP (if those stem from binary systems at all). This interpretation is consistent with the photometric and spectroscopic similarities of SN~2013ej to Type~IIP SNe \citep[e.g.,][]{Bose2015, Huang2015, Dhungana2016, Valenti2016, Yuan2016}.

In contrast to SN~2013ej and Type~IIP SNe, SN~2017ahn showed a low degree of polarization not only during the photospheric phase but also in the radioactive decay tail phase. As discussed in \S 4.2, this can be taken to mean that the explosion geometry of SN~2017ahn was relatively spherical although there is a finite probability that the weak polarization is due to a projection effect. If the explosion of SN~2017ahn was genuinely spherical, its progenitor should be intrinsically different from those of SN~2013ej and Type~IIP SNe.  In that case, the steeper decline of the LC of SN~2017ahn compared to Type~IIP SNe could be attributed either to the smaller hydrogen envelope caused by a stronger stellar-wind mass loss or to an interaction with a radially confined CSM created by some unidentified stellar instabilities not long before the SN explosion. However, if SN~2017ahn underwent a genuinely aspherical explosion (i.e., the polarization was low because of a projection effect), the main origin of the fast LC decline should be similar as in SN~2013ej, i.e., either a small hydrogen envelope or CSM interaction, resulting from interaction in a binary system.

Based on a comparison of early photometric and spectroscopic observations with hydrodynamical and radiative transfer models, \citet[][]{Tartaglia2021} suggested that the progenitor of SN~2017ahn was a massive yellow super/hypergiant deprived of most of its hydrogen envelope. This conclusion lends support to a scenario in which the progenitor of SN~2017ahn was a massive star with an initial mass larger than what is typical of Type~IIP SNe. Its hydrogen envelope was stripped off by enhanced mass loss before the star was eventually destroyed in a spherically symmetric explosion.

Even if SN~2017ahn did share the explosion geometry with SN~2013ej (and possibly other Type~II SNe), it is unlikely that it had the same aspherical CSM interaction as SN~2013ej had. In that case, we would need to have been looking at SN~2017ahn along the polar axis of the aspherical CSM (because SN~2017ahn presented low polarization also in the photospheric phase).  However, this requires an extremely ad-hoc scenario.  The polar axis of the aspherical CSM would need to be identical to the polar axis of the aspherical explosion and both axes would need to have the same inclination angle with respect to the line of sight:  we discard this as a valid hypothesis. From the different polarimetric properties of SN~2013ej and SN~2017ahn, we conclude that Type~IIL SNe occur in progenitors that have different explosion properties and/or different pre-explosion mass-loss processes.

Finally, we note that our polarimetric data for Type~IIL SNe do not exclude a different interpretation. If Type~IIP SNe and the Type~IIL SN~2013ej have qualitatively similar, very aspherical explosion geometries, there should be events that are observed down the polar axis so that they do not exhibit a net polarization signal.  With only about a handful of well enough observed events, the absence of pole-on events is not in conflict with the common-geometry hypothesis.  However, SN~2017ahn could fill the gap without causing any statistical problem for the combined IIL/P sample. If this conjecture has any basis, the steepness of the IIL LCs would still have more than one origin, namely either a smaller hydrogen envelope or an interaction with radially confined CSM or both.

\section{Conclusions}

We have presented and discussed spectropolarimetry of two Type~IIL SNe: SN~2013ej and SN~2017ahn, to the best of our knowledge the only Type~IIL SN candidates that have polarimetric coverage from the photospheric through the tail phases. SN~2013ej showed a high polarization degree from the very beginning of the photospheric phase to the tail phase with time-variable polarization angles, indicating two components, namely an interaction with aspherical CSM and an aspherical explosion similar to the explosions of Type~IIP SNe. This implies that SN~2013ej originated from a progenitor analogous to those of Type~IIP SNe, with the notable difference that the explosion was accompanied by an aspherical CSM interaction.  In contrast, SN~2017ahn was characterized by a low polarization from the photospheric phase to the tail phase, indicating a relatively spherical explosion. From the differences between SN~2013ej and SN~2017ahn, we conclude that Type~IIL SNe have, at least, two different origins: they originate from progenitors that have different explosion properties and/or different mass-loss processes. In this scenario, 2013ej-like Type~IIL SNe would have a progenitor similar to those of Type~IIP SNe, with the ejecta undergoing an aspherical CSM interaction. 2017ahn-like Type~IIL SNe would instead come from a more massive progenitor that was depleted of most of its hydrogen envelope and disrupted by a much more spherical explosion.

The sample of Type~IIL events covered by spectropolarimetry is still very poor. In view of the powerful diagnostics provided by this technique, it is very important to increase the amount of data, which is key to understanding the origin and diversity of the LC shapes of these objects.

\vspace{\baselineskip}
\noindent
{\bf Acknowledgements.} This paper is based on observations made with ESO Telescopes at the La Silla Paranal Observatory under program IDs 091.D-0401 and 098.D-0852 and data obtained from the ESO Science Archive Facility and the Berkeley SuperNova DataBase. The authors are grateful to ESO's Paranal staff for the support given during the service mode observations of SN~2017ahn. 
T.N. was supported by a Japan Society for the Promotion of Science (JSPS) Overseas Research Fellowship and is funded by the Academy of Finland project 328898.
Time domain research by D.J.S. is supported by NSF grants AST-1821987, 1813466, \& 1908972, and by the Heising-Simons Foundation under grant \#2020-1864.
M.B. acknowledges support from the Swedish Research Council (Reg. no. 2020-03330).
H.K. was funded by the Academy of Finland projects 324504 and 328898.
L.T. acknowledges support from MIUR (PRIN 2017 grant 20179ZF5KS).

\vspace{\baselineskip}
\noindent
{\bf Data availability.}
The data underlying this article are available in the ESO Science Archive Facility at http://archive.eso.org.





\appendix

\section{The complete spectropolarimetric data set}

\begin{figure*}
\includegraphics[width=1.8\columnwidth]{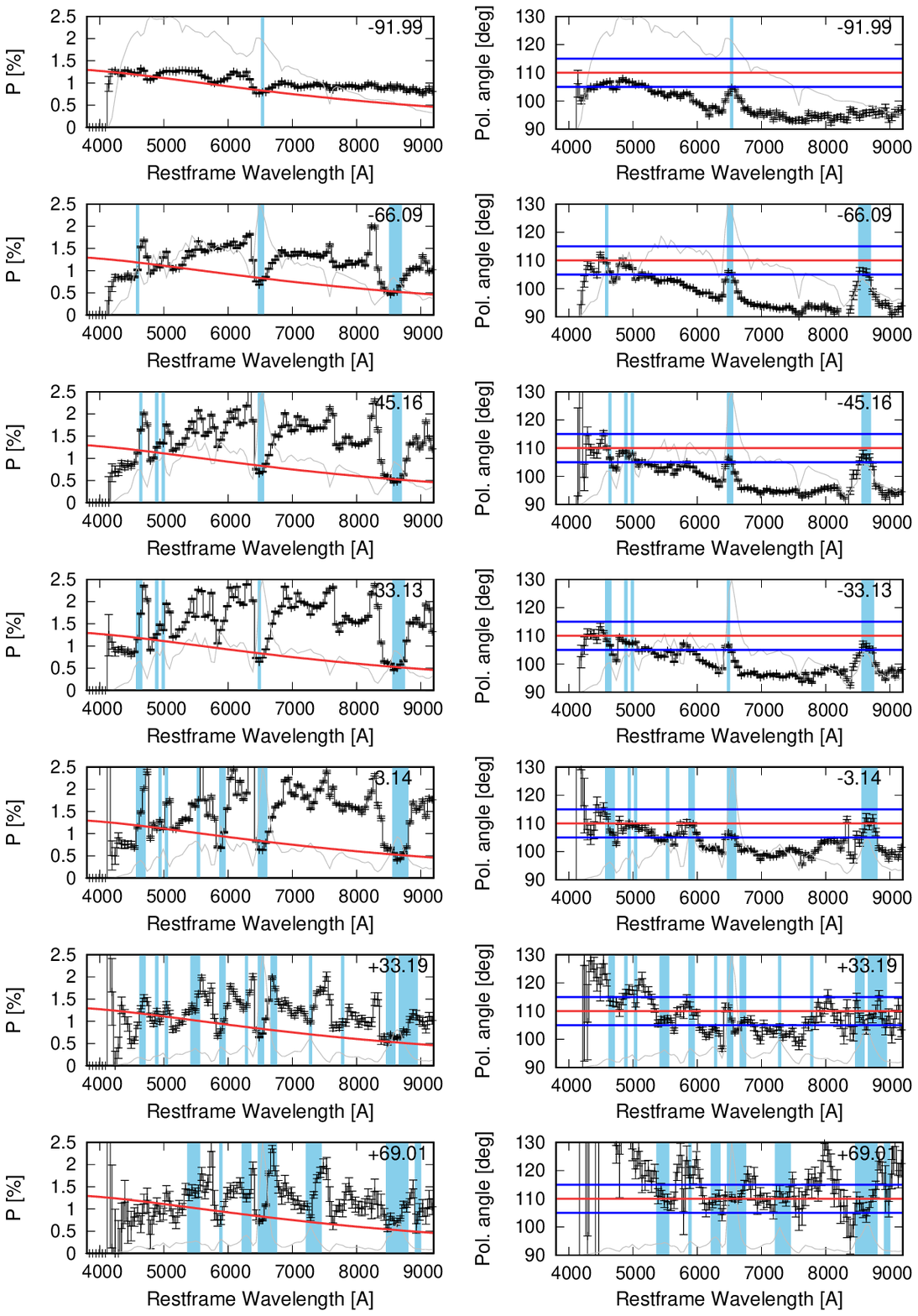}
    \caption{Polarization degree and angle of SN~2013ej before ISP subtraction. Each frame is labeled with the phase of the observations. The data are binned to $50$ Angstrom per point. The gray lines in the background of each plot are the unbinned flux spectra at the same epochs. The ISP is described by $P(\lambda) = P_{\rm{max}} \exp \left[ -K \ln^{2} \left( \lambda_{\rm{max}}/\lambda \right) \right]$, where $P_{\rm{max}}=1.32$, $\lambda_{\rm{max}}=3300$ Angstrom and $K=1.0$ (red lines). The red line and the blue lines in the polarization angle plot represent the assumed ISP angle ($\theta=110$ degrees) and the adopted maximum and minimum polarization angle ($\theta=105$ and $\theta=115$ degrees), respectively, for determining the ISP. The blue hatching shows the wavelength ranges used for the ISP estimate.
    }
    \label{figa1}
\end{figure*}

\begin{figure*}
\includegraphics[width=1.8\columnwidth]{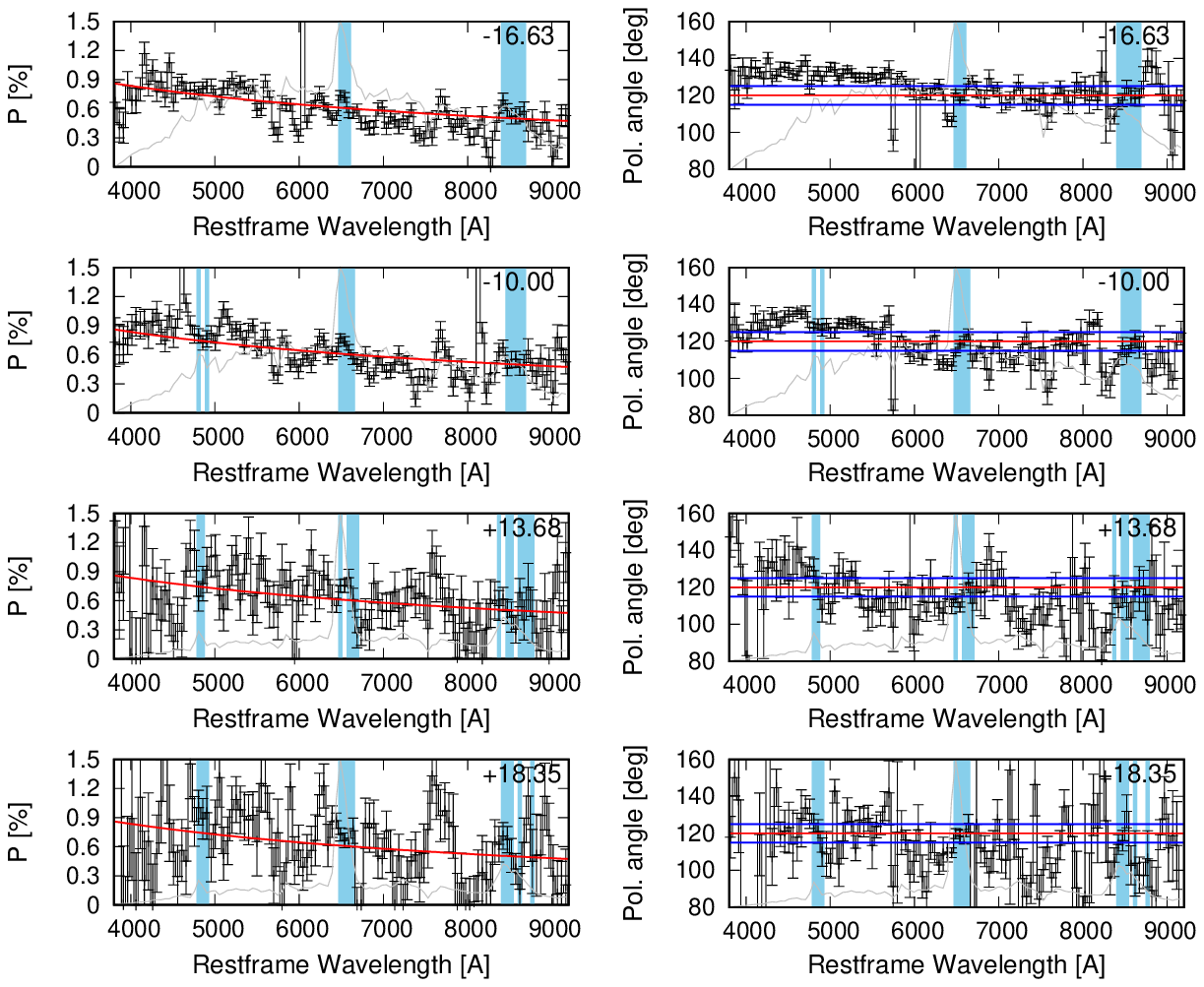}
    \caption{Same as Fig.~\ref{figa1}, but for SN~2017ahn. The red lines shows the case with $P_{\rm{max}}=2.05$, $\lambda_{\rm{max}}=200$ Angstrom and $K=0.1$. Here, the assumed ISP angle is $\theta=120$ degrees. 
    }
    \label{figa2}
\end{figure*}

\begin{figure*}
\includegraphics[width=0.9\columnwidth]{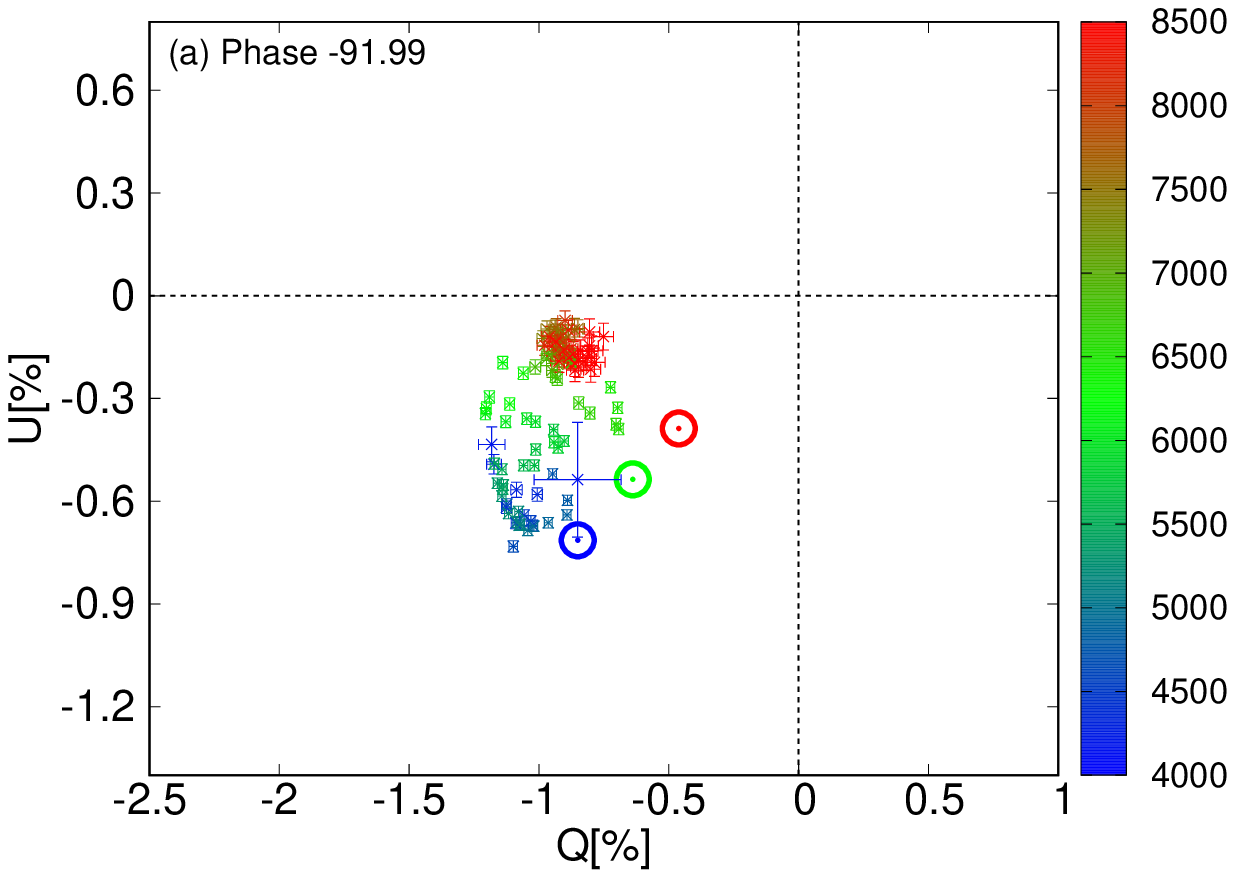}
\includegraphics[width=0.9\columnwidth]{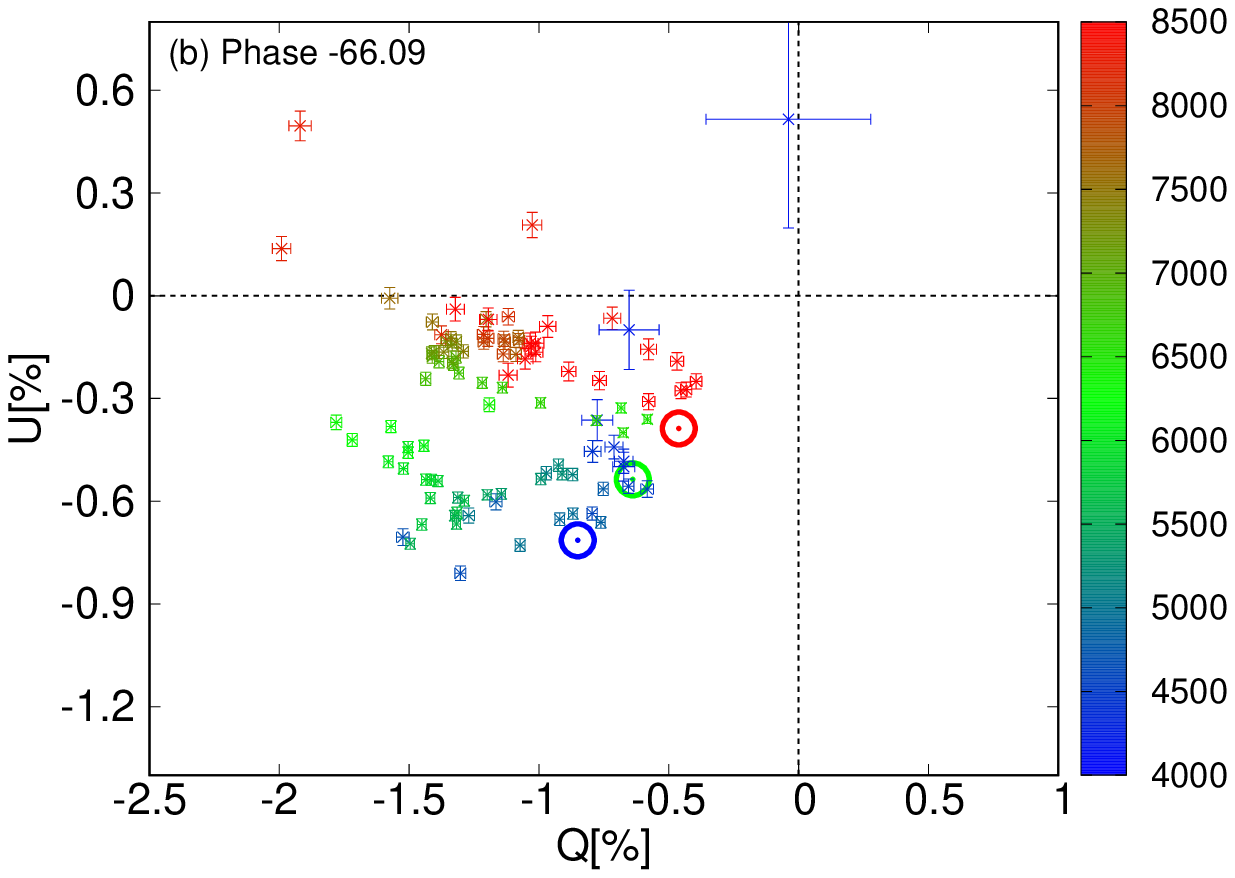}
\includegraphics[width=0.9\columnwidth]{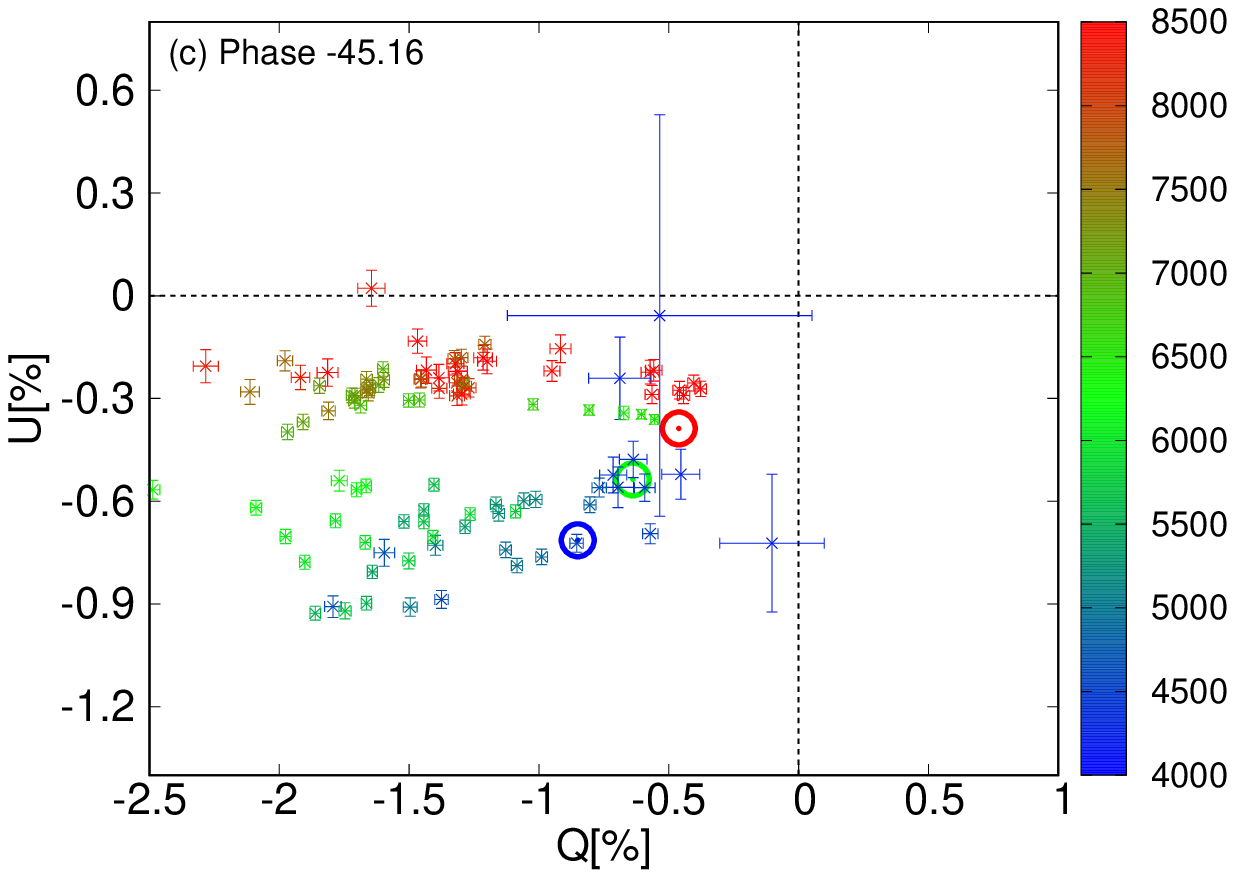}
\includegraphics[width=0.9\columnwidth]{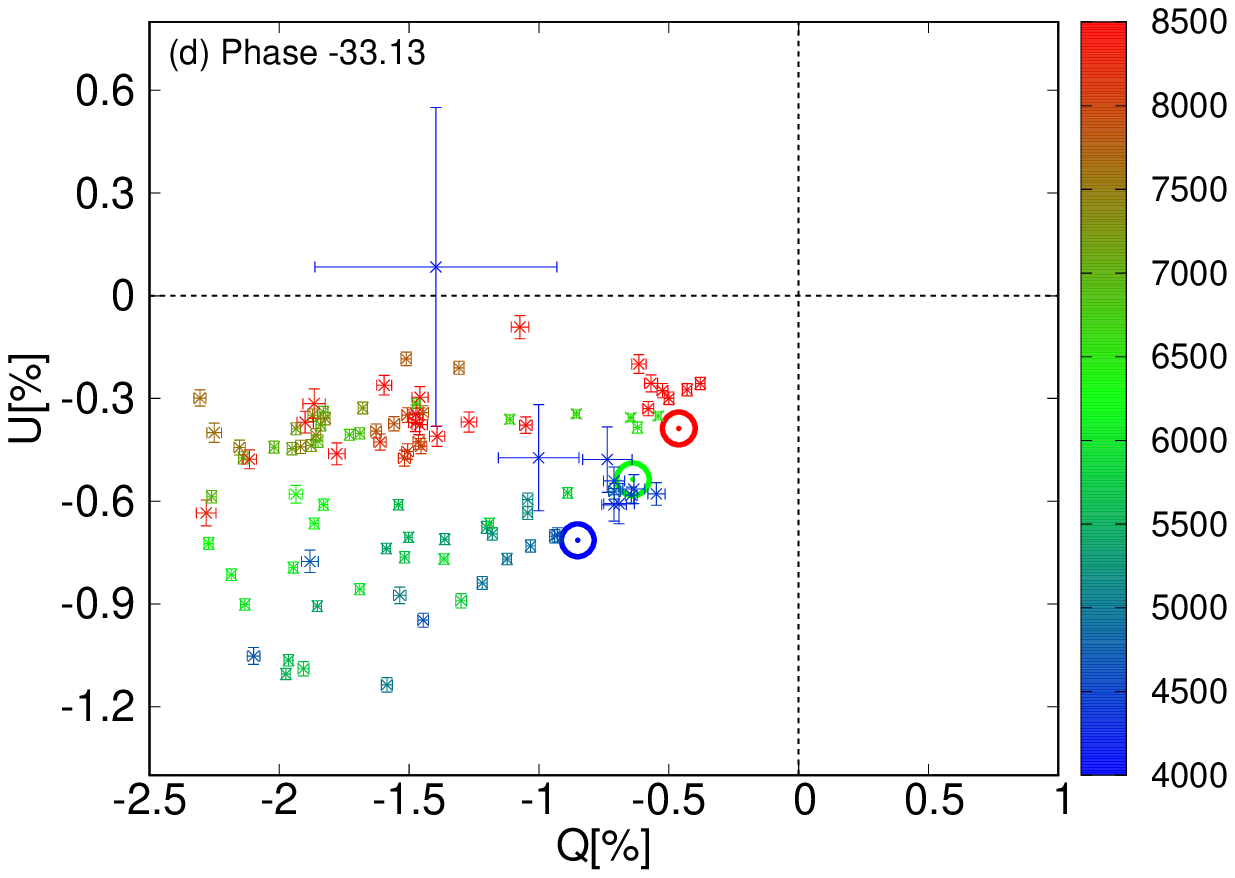}
\includegraphics[width=0.9\columnwidth]{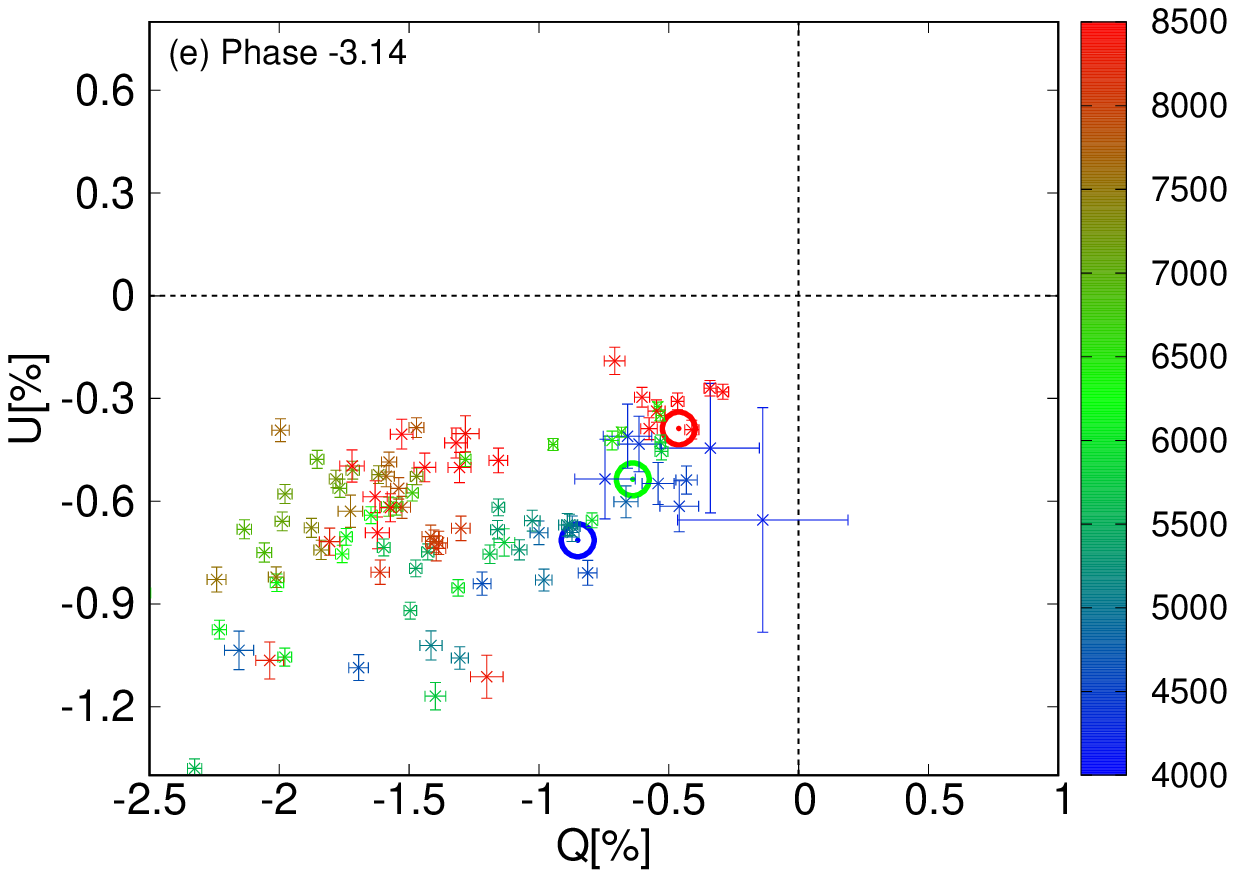}
\includegraphics[width=0.9\columnwidth]{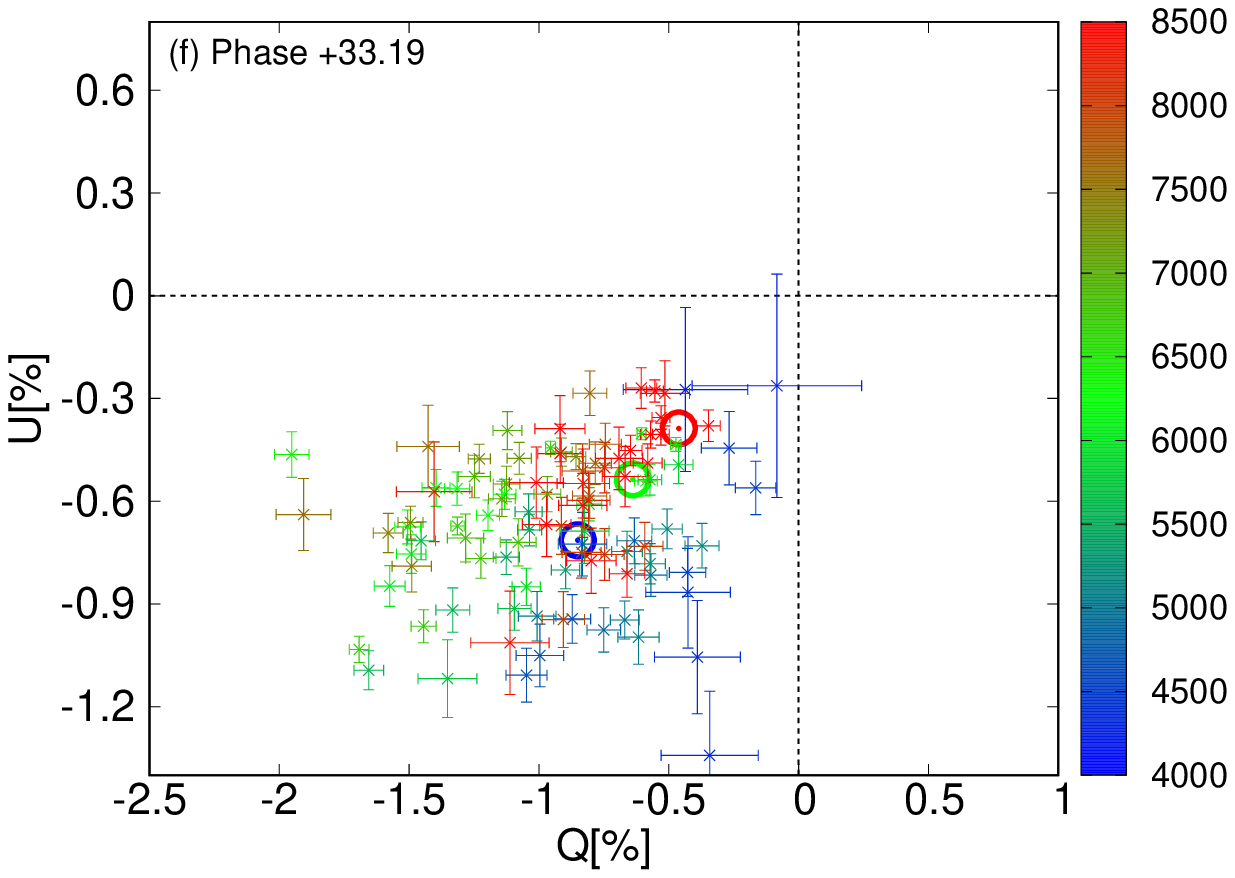}
\includegraphics[width=0.9\columnwidth]{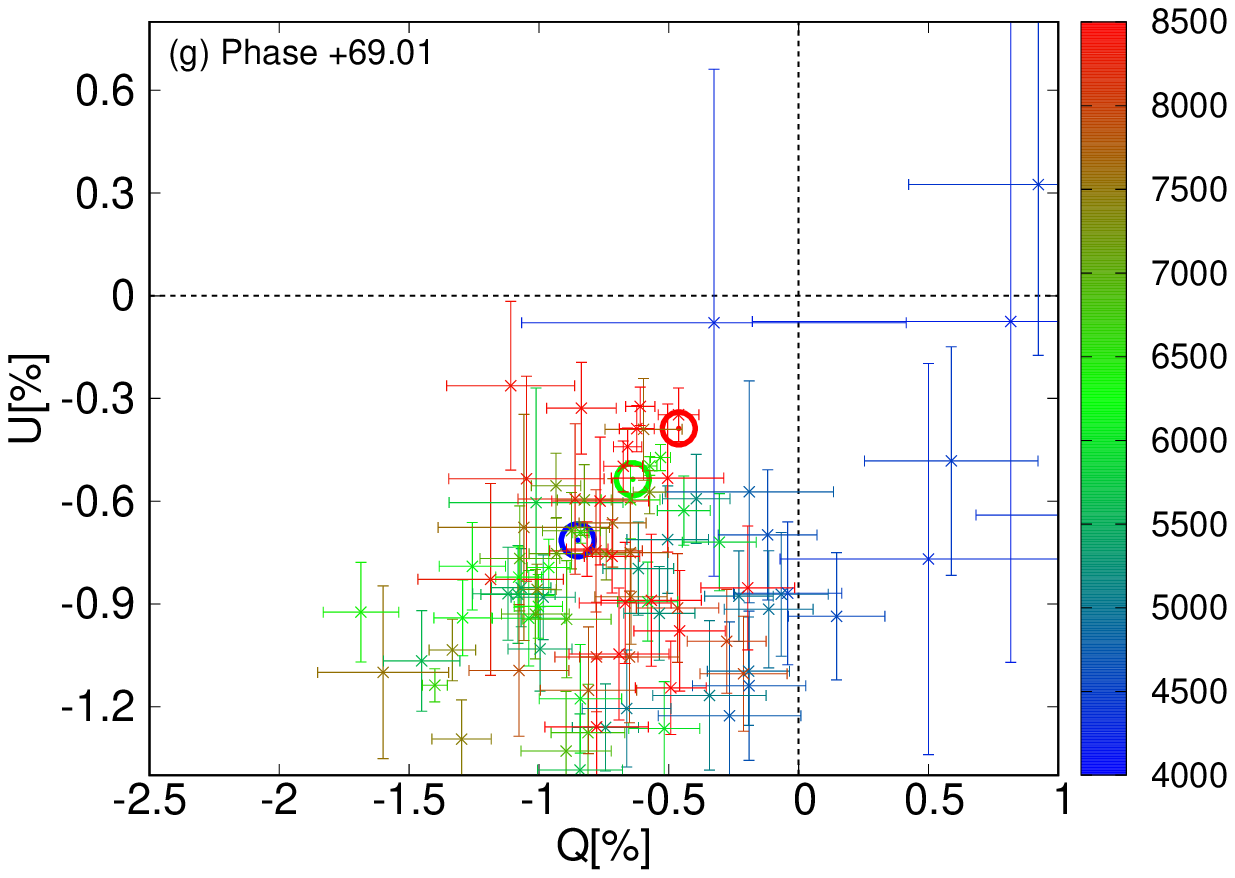}
    \caption{Spectropolarimetric data of SN~2013ej in the Q-U plane before ISP subtraction. The estimated ISP at $\lambda=5000, 6500$ and 8000 Angstrom is shown with the blue, green and red circles, respectively. The colors encode the wavelengths (Angstrom) as shown at the right edge of each panel.
    }
    \label{figa3}
\end{figure*}

\begin{figure*}
\includegraphics[width=0.9\columnwidth]{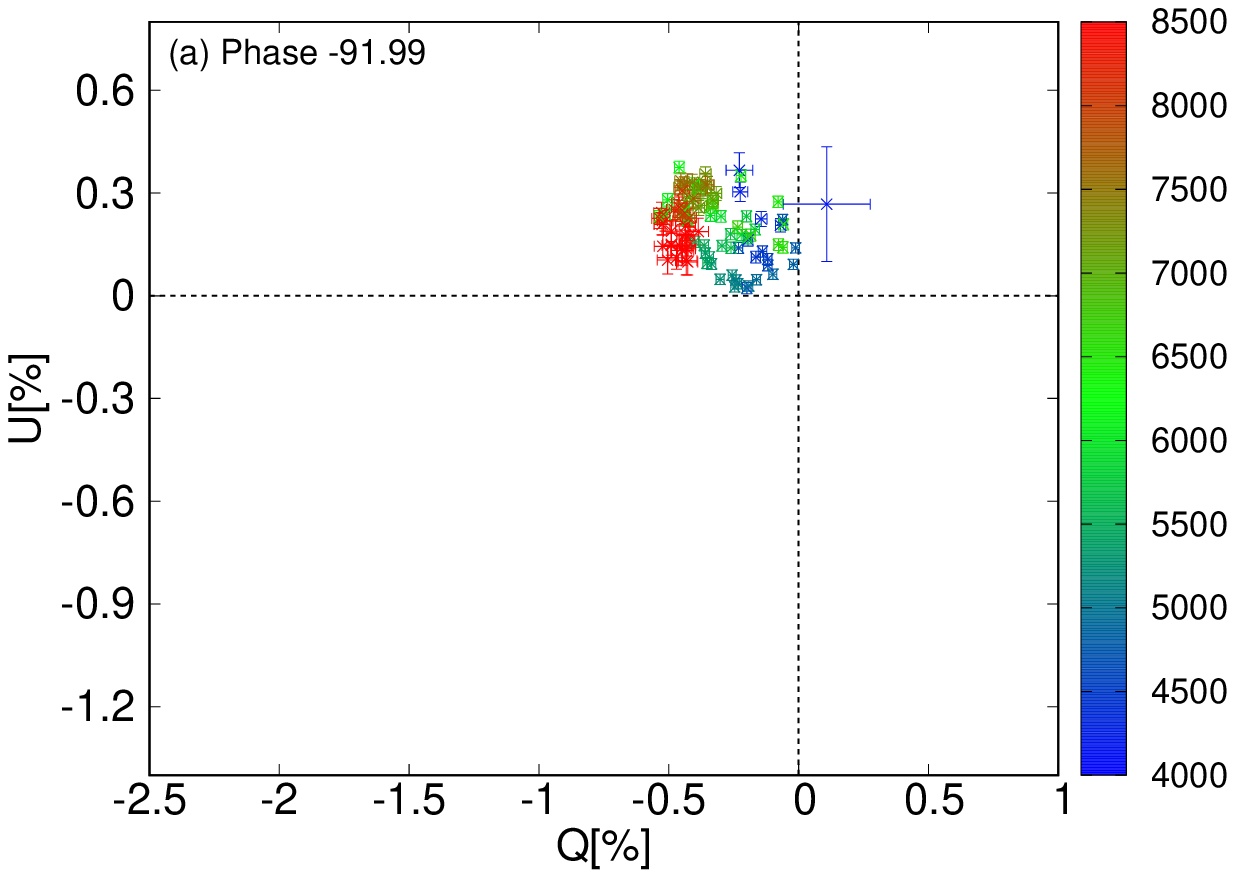}
\includegraphics[width=0.9\columnwidth]{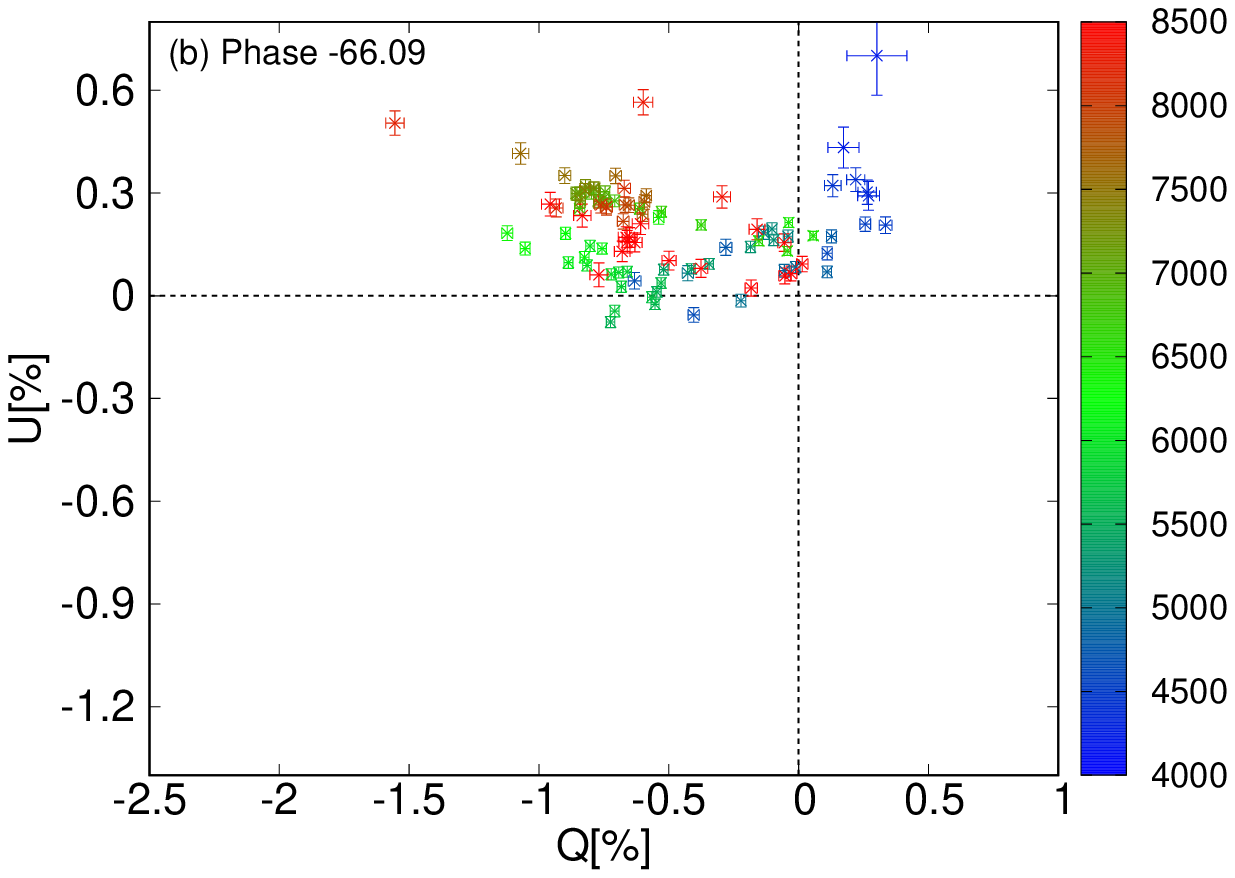}
\includegraphics[width=0.9\columnwidth]{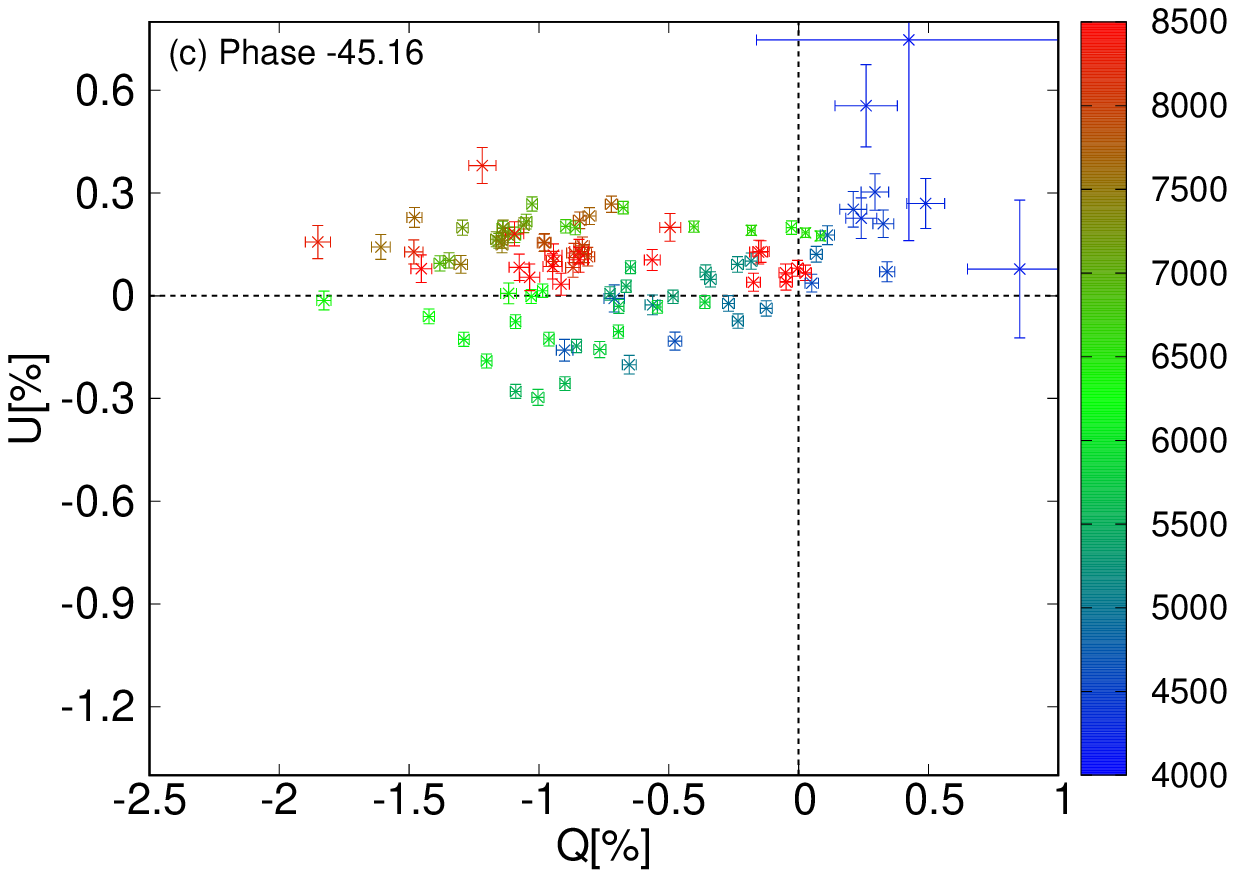}
\includegraphics[width=0.9\columnwidth]{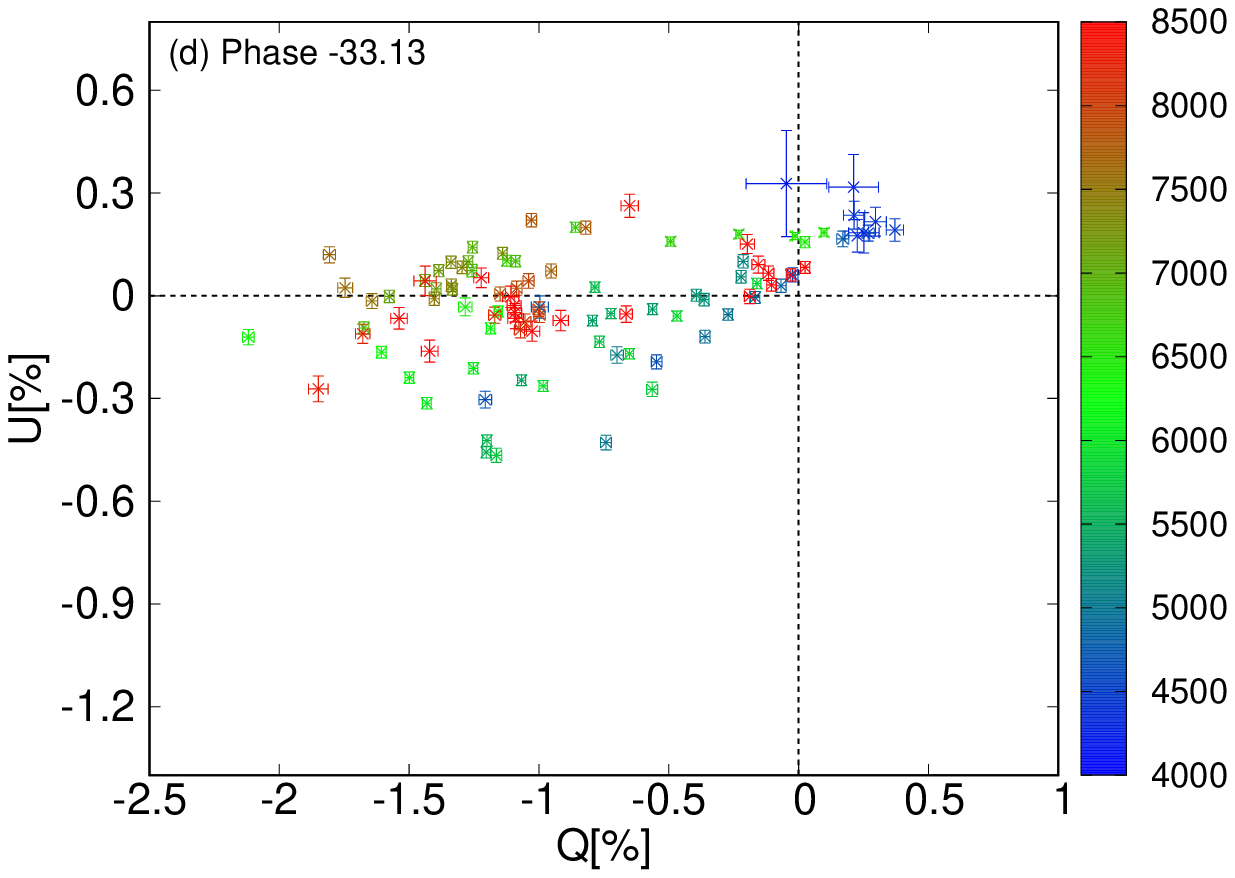}
\includegraphics[width=0.9\columnwidth]{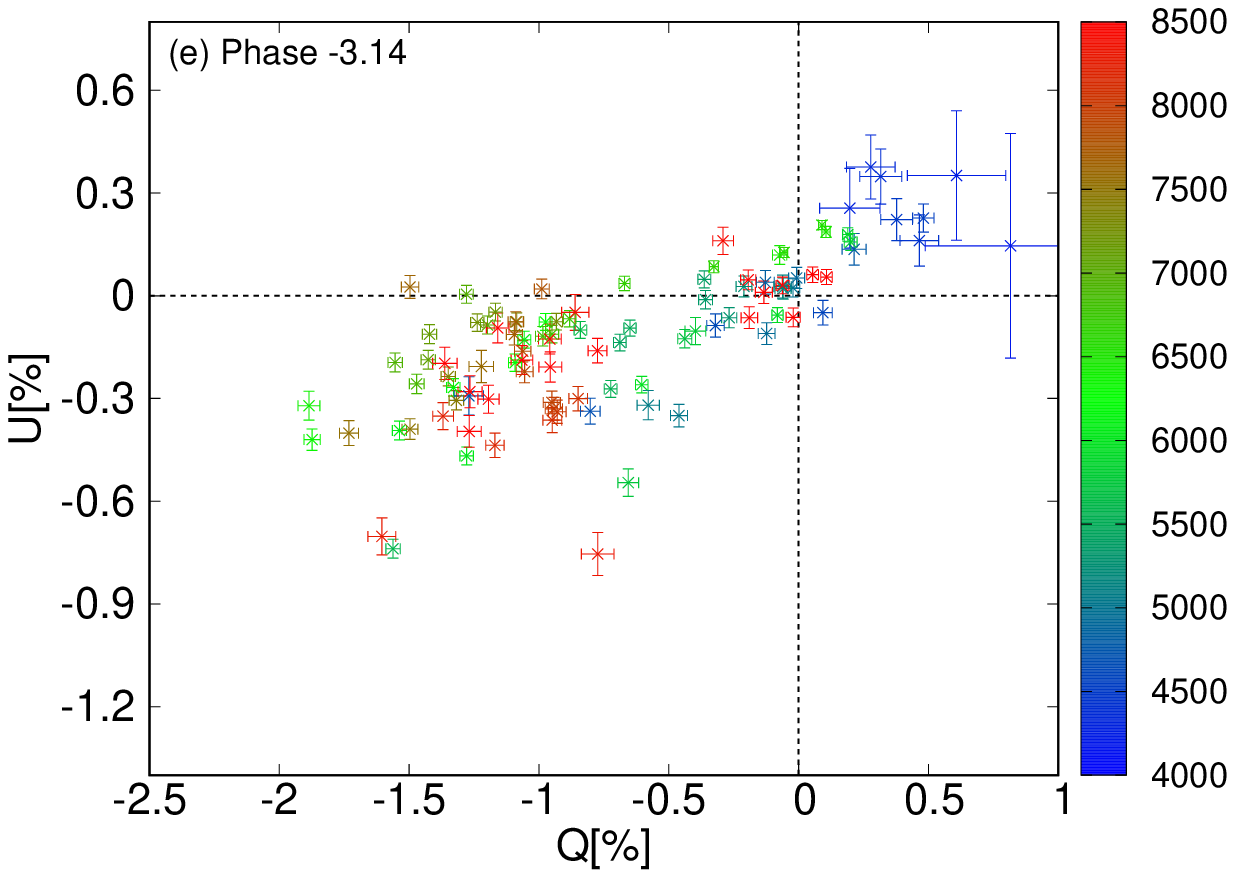}
\includegraphics[width=0.9\columnwidth]{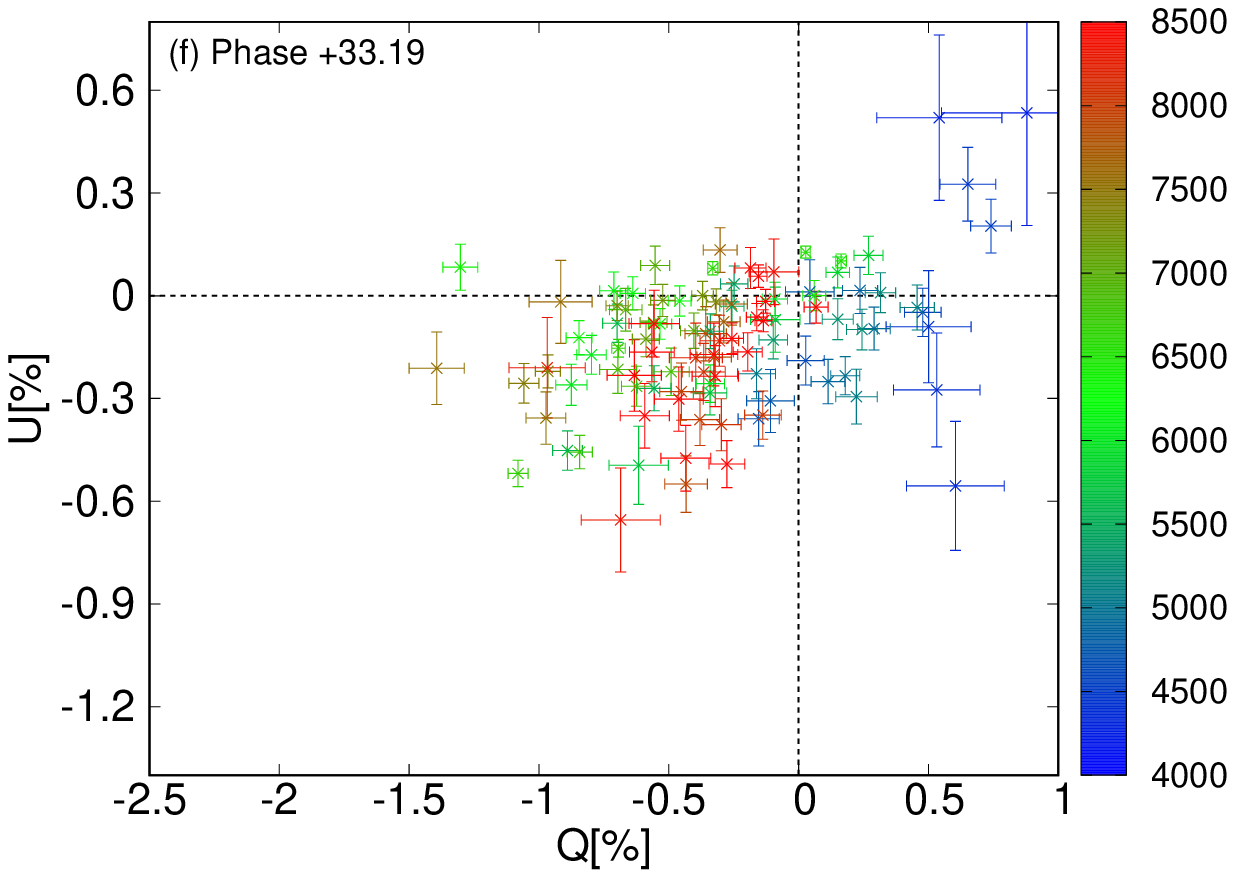}
\includegraphics[width=0.9\columnwidth]{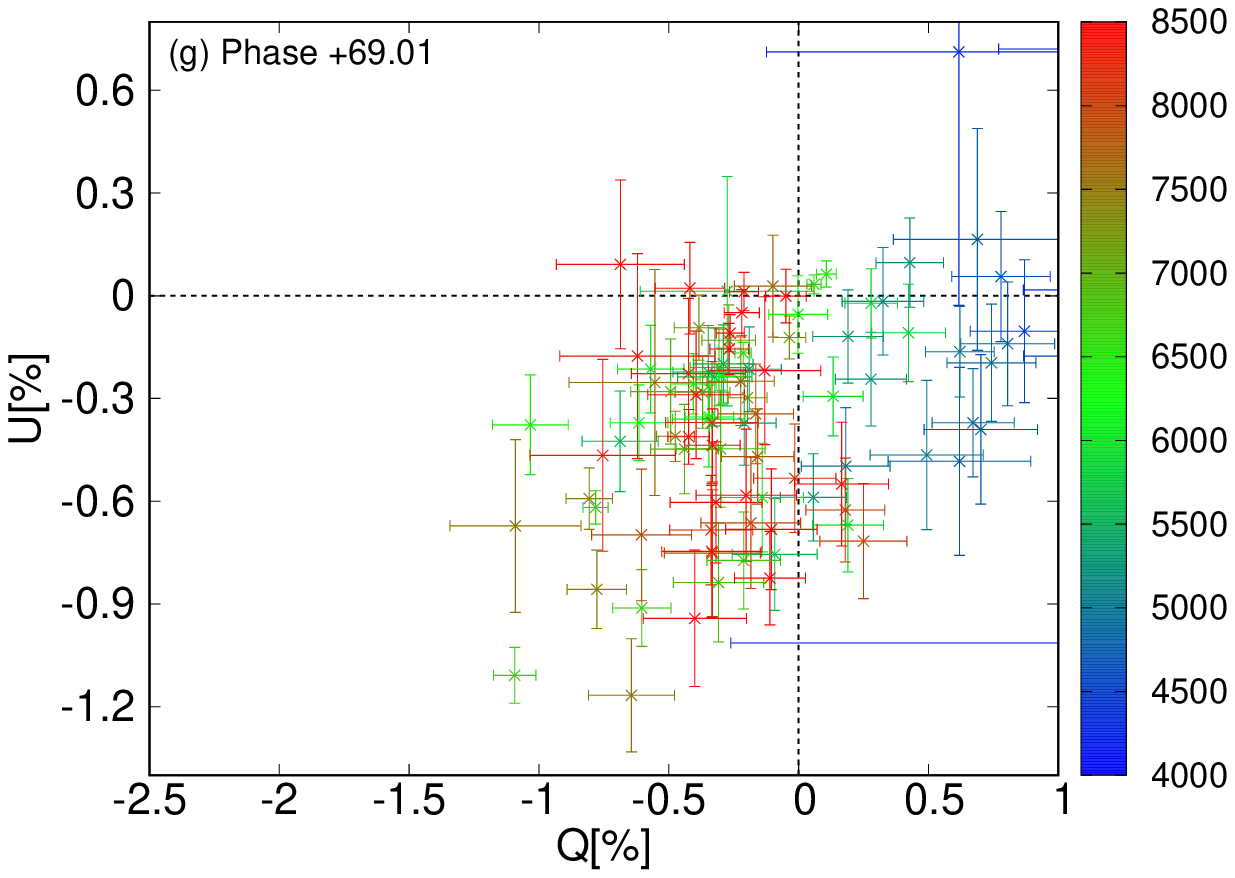}
    \caption{Same as Fig.~\ref{figa3}, but for the data after ISP subtraction.
    }
    \label{figa4}
\end{figure*}

\begin{figure*}
\includegraphics[width=0.9\columnwidth]{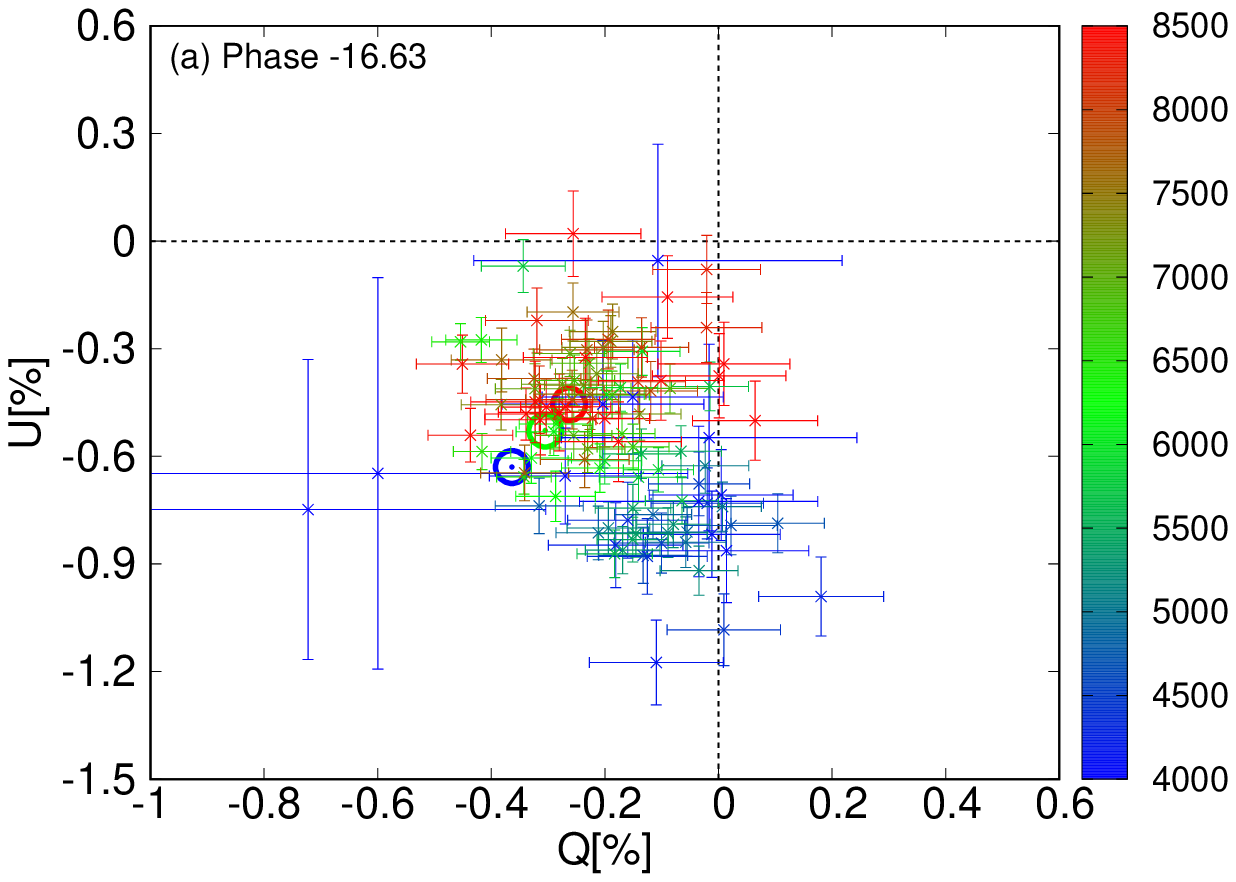}
\includegraphics[width=0.9\columnwidth]{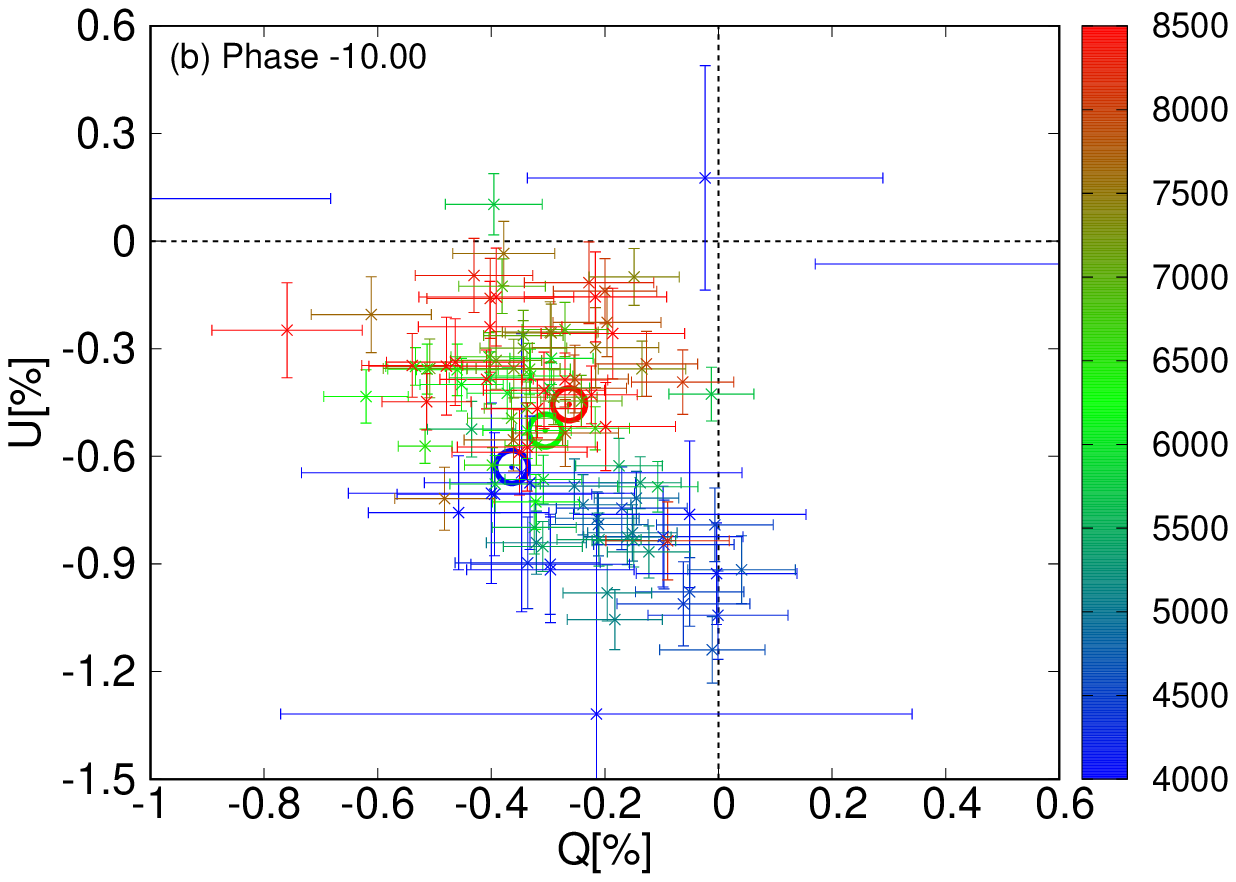}
\includegraphics[width=0.9\columnwidth]{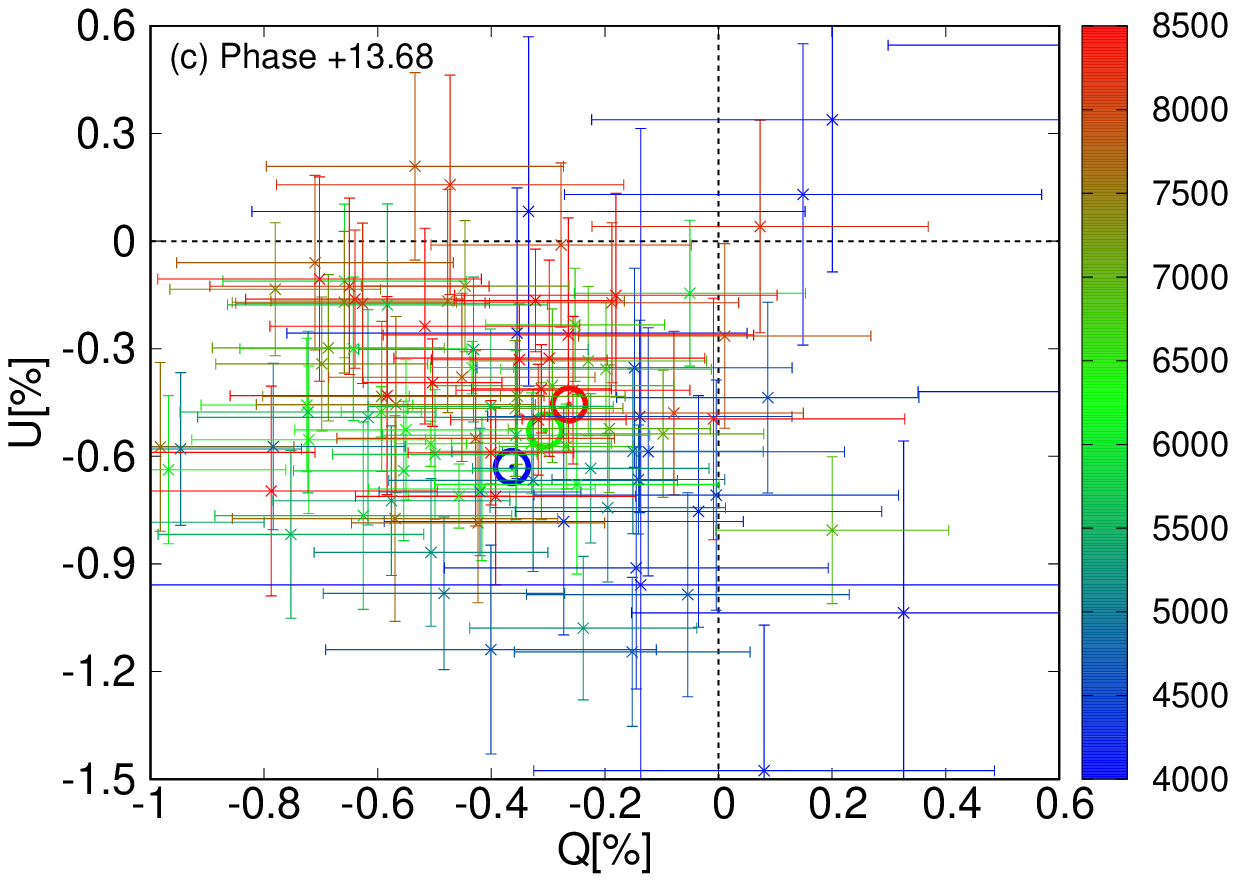}
\includegraphics[width=0.9\columnwidth]{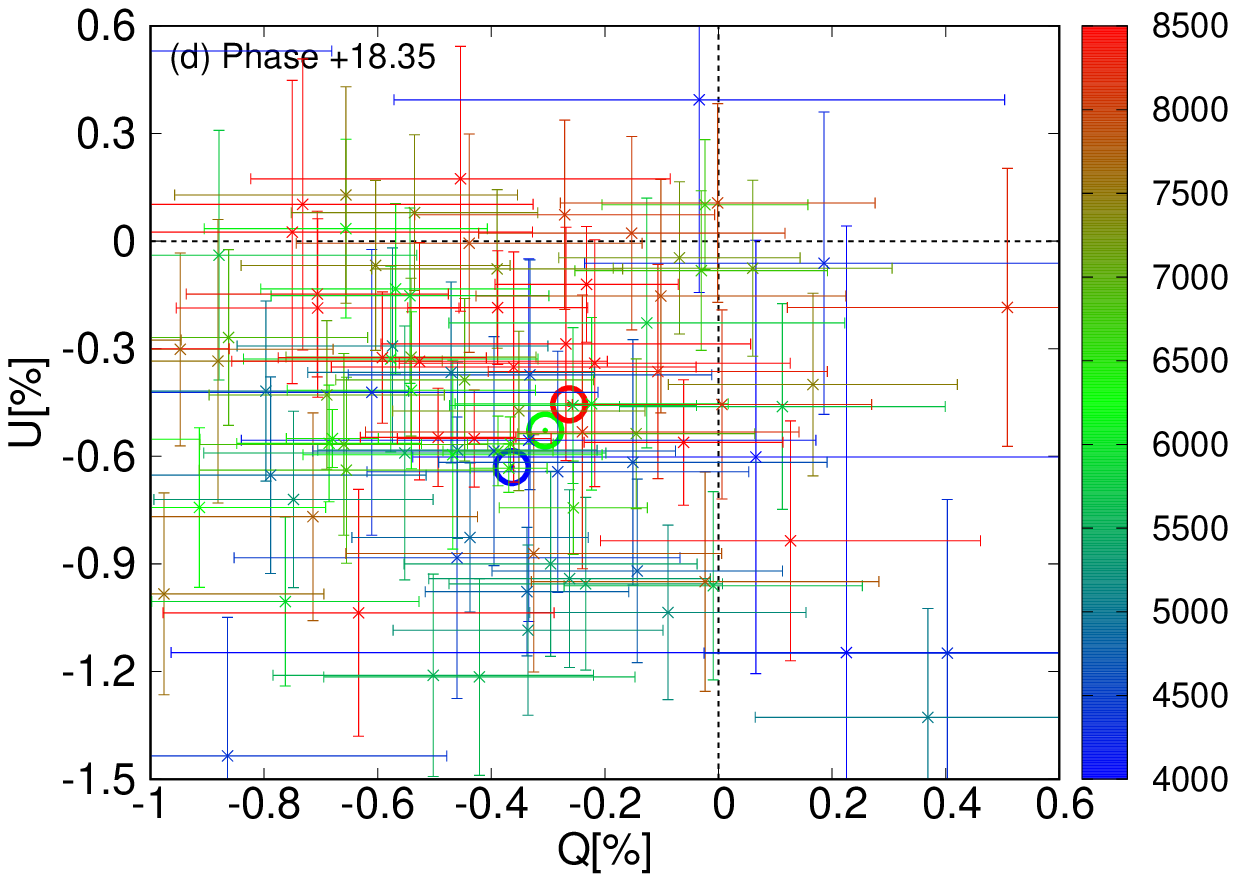}
    \caption{Same as Fig.~\ref{figa3}, but for SN~2017ahn.
    }
    \label{figa5}
\end{figure*}

\begin{figure*}
\includegraphics[width=0.9\columnwidth]{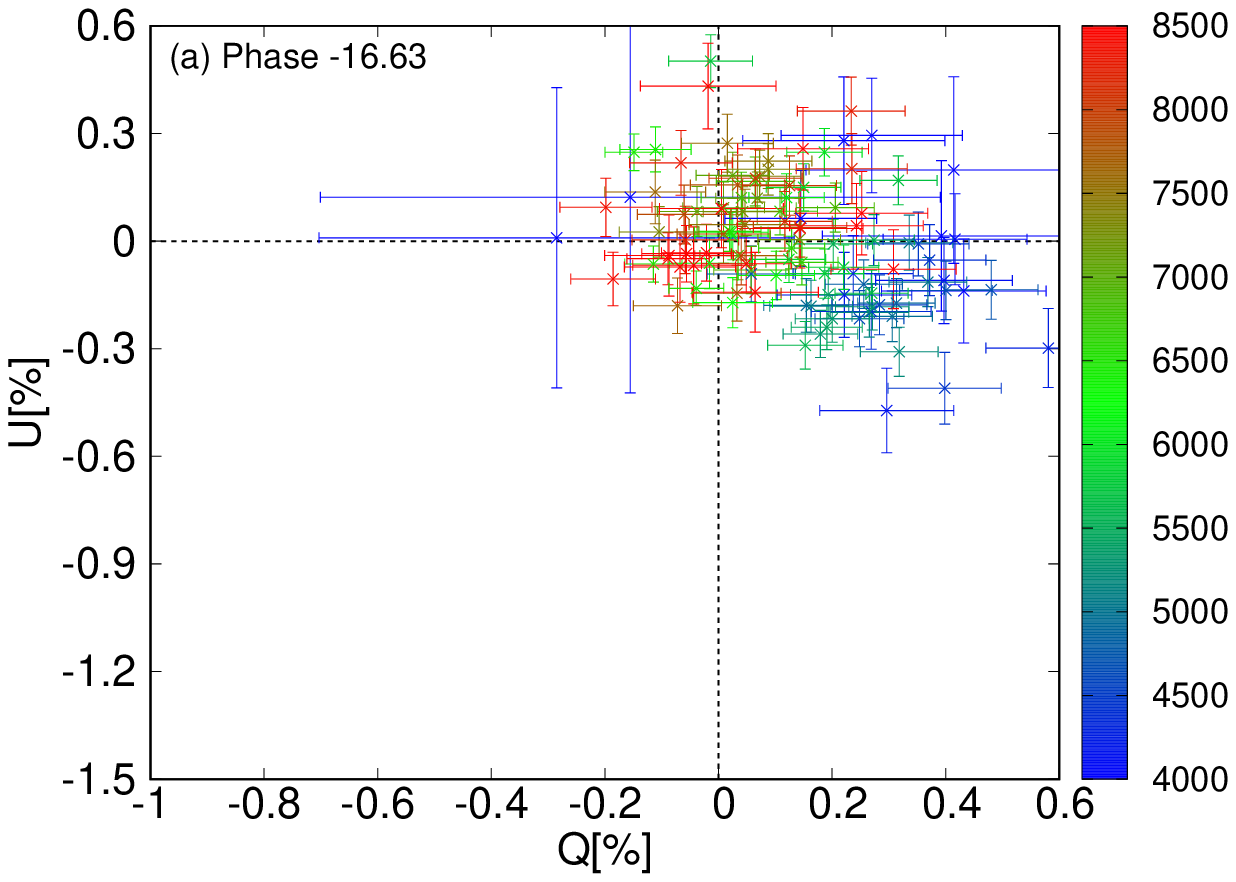}
\includegraphics[width=0.9\columnwidth]{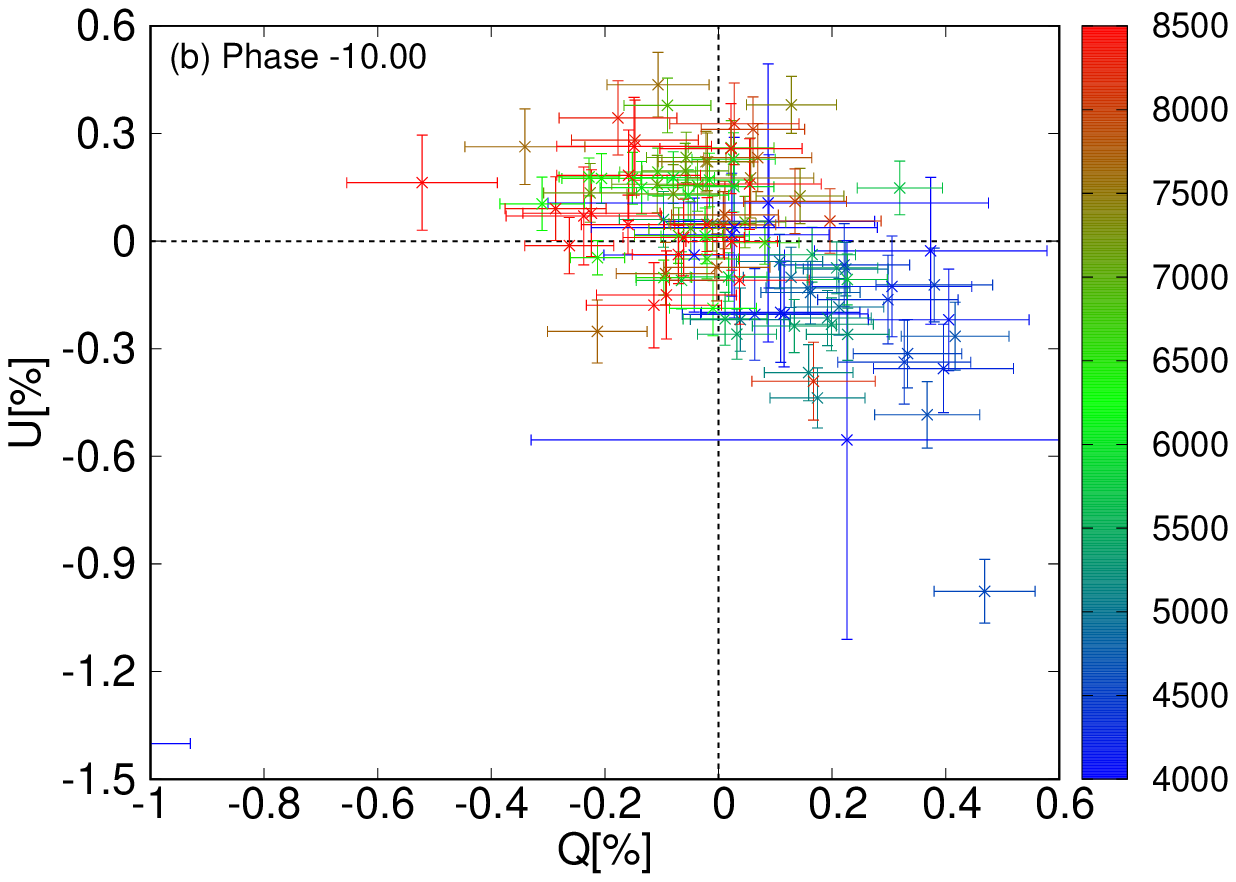}
\includegraphics[width=0.9\columnwidth]{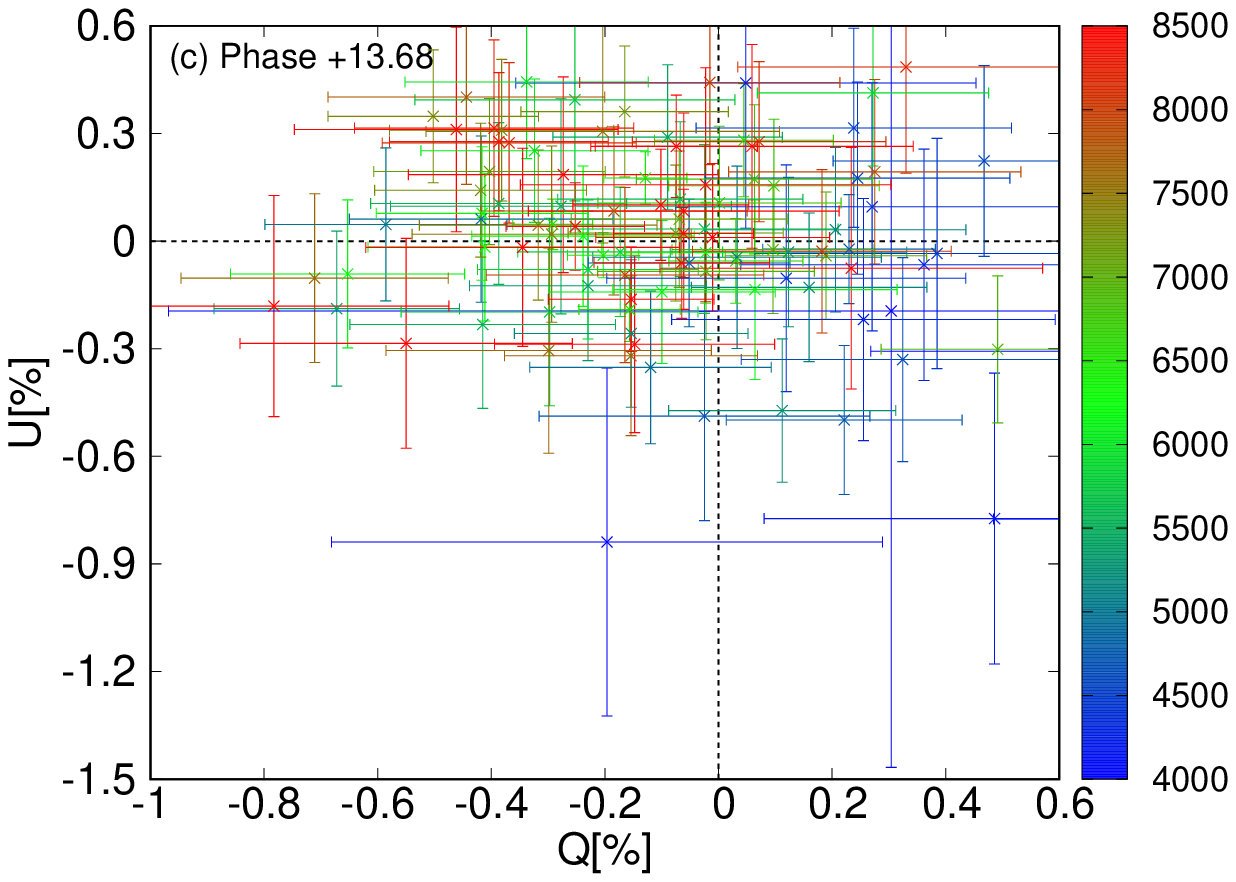}
\includegraphics[width=0.9\columnwidth]{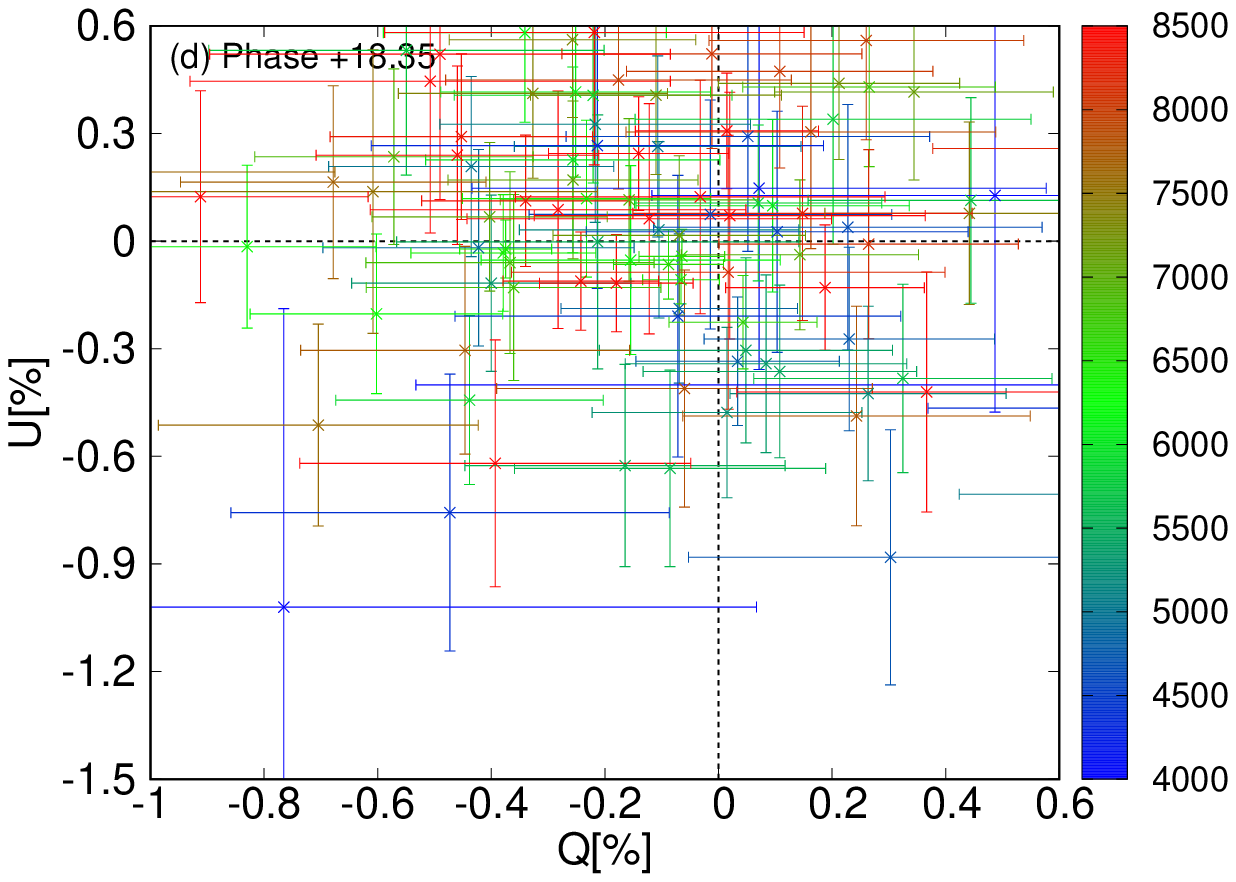}
    \caption{Same as Fig.~\ref{figa5}, but for the data after ISP subtraction.
    }
    \label{figa6}
\end{figure*}

In Figs.~\ref{figa1} to \ref{figa6}, we provide all epochs of  spectropolarimetric data for SN~2013ej and SN~2017ahn. Figs.~\ref{figa1} and \ref{figa2} show the observed polarization degree and angle of SN~2013ej and SN~2017ahn, respectively, before ISP subtraction. Figs.~\ref{figa3} and \ref{figa4} present the data on the Stokes Q-U plane for SN~2013ej before and after ISP subtraction, respectively, while Figs.~\ref{figa5} and \ref{figa6} are the same but for SN~2017ahn.

The data in the Q-U diagrams of SN~2013ej show a displacement from the origin  (Fig.~\ref{figa3}), indicating the presence of an ISP component. In addition, the estimated ISP in this study is located at the edge of the cloud of the data points, which is the proper place for ISP in the Q-U plane. The ISP-subtracted data spread along a dominant axis from the origin, which is another manifestation of an aspherical structure with a dominant symmetry axis (Fig.~\ref{figa4}). For SN~2017ahn, Fig.~\ref{figa5} indicates an ISP component in the Q-U plane that is likewise consistent with our ISP estimate. After ISP correction, the data cluster around the origin in Fig.~\ref{figa6}, demonstrating that the intrinsic polarization of SN~2017ahn is low. These alternative presentations of the data confirm for both SNe the adequacy of our method to determine the ISP components.

\bsp	
\label{lastpage}
\end{document}